\documentclass[a4paper,11pt]{article}
\pdfoutput=1
\usepackage{jcappub}

\usepackage{color}
\usepackage{amsfonts}
\usepackage{graphicx}
\usepackage{amssymb}
\usepackage{amsmath}

\usepackage{eufrak}
\usepackage{mathrsfs}
\def\psiP{\mathscr{P}}

\renewcommand{\thesection}{\arabic{section}}

\def\laq{~\raise 0.4ex\hbox{$<$}\kern -0.8em\lower 0.62ex\hbox{$\sim$}~}
\def\gaq{~\raise 0.4ex\hbox{$>$}\kern -0.7em\lower 0.62ex\hbox{$\sim$}~}
\def \g#1#2{\int_{#1}^{\eta_o}d#2\psi(#2)}

\def \g#1#2{\int_{#1}^{\eta_o}d#2\psi(#2)}
\def \G#1#2{\psi(#1)+2\int_{#1}^{\eta_o}d#2\pa_{{#2}}\psi(#2)}

\def\beq{\begin{equation}}
\def\eeq{\end{equation}}
\def\bea{\begin{eqnarray}}
\def\eea{\end{eqnarray}}
\def\bean{\begin{eqnarray*}}
\def\eean{\end{eqnarray*}}

\def\vp{\varphi}

\def \pa {\partial}

\def \ti {\widetilde}
\def \la {\lambda}

\def \b {\beta}
\def \a {\alpha}

\def \ga {\gamma}
\def \sg {\sigma}

\def \ep {\epsilon}
\def \r {\rho}

\def \ep {\eta^{+}}
\def \tt {\ti\theta}
\def \tta {\ti\theta^a}
\def \U{\Upsilon}
\def \pat {\pa}

\def\th#1#2{\theta^{(#1)#2}}
\def\Ep#1{\eta^{(#1)+}}

\usepackage[english]{babel}
\usepackage[utf8]{inputenc}
\usepackage{amsmath}
\usepackage{color}

\def\laq{~\raise 0.4ex\hbox{$<$}\kern -0.8em\lower 0.62ex\hbox{$\sim$}~}
\def\gaq{~\raise 0.4ex\hbox{$>$}\kern -0.7em\lower 0.62ex\hbox{$\sim$}~}

\def\be{\begin{equation}}
\def\ee{\end{equation}}
\def \ga {\gamma}

\def\beq{\begin{equation}}
\def\eeq{\end{equation}}
\def\bea{\begin{eqnarray}}
\def\eea{\end{eqnarray}}

\def \pa {\partial}

\def \ti {\widetilde}

\newcommand{\Ups}{\Upsilon}

\newcommand{\Acal}{\mathcal A}

\newcommand{\Hcal}{\mathcal H}

\def\laq{~\raise 0.4ex\hbox{$<$}\kern -0.8em\lower 0.62ex\hbox{$\sim$}~}
\def\gaq{~\raise 0.4ex\hbox{$>$}\kern -0.7em\lower 0.62ex\hbox{$\sim$}~}

\def\beq{\begin{equation}}
\def\eeq{\end{equation}}
\def\bea{\begin{eqnarray}}
\def\eea{\end{eqnarray}}
\def\bean{\begin{eqnarray*}}
\def\eean{\end{eqnarray*}}

\def\vp{\varphi}

\def \pa {\partial}

\def \ti {\widetilde}
\def \la {\lambda}

\def \b {\beta}
\def \a {\alpha}

\def \ga {\gamma}
\def \sg {\sigma}

\def \ep {\epsilon}
\def \r {\rho}

\def \ro {\left( \eta^+-\eta \right)^{(0)}}
\def \ep {\eta^{+}}
\def \tt {\ti\theta}

\def \U{\Upsilon}

\def\th#1#2{\theta^{#2(#1)}}
\def\Ep#1{\eta^{(#1)+}}

\def \r#1#2{\frac{#1-#2}{\eta_o-#1}}

\def \g#1#2{\int_{#1}^{\eta_o}d#2\psi(#2)}
\def \r#1#2{\frac{#1-#2}{\eta_o-#1}}
\def \ro{\Ep{0}-\eta^{(0)}}

\def\th#1#2{\theta^{#2(#1)}}

\def\Ep#1{\eta^{+(#1)}}

\title{ A new  approach to the propagation of light-like signals in perturbed cosmological backgrounds}

\author[a,b]{G. Fanizza}
\author[a]{, M. Gasperini}
\author[b]{, G. Marozzi}
\author[c]{, G. Veneziano}

\affiliation[a]{Dipartimento di Fisica, Universit\`a di Bari, \\ 
Via G. Amendola 173, 70126 Bari, Italy}
\affiliation[b]{
Universit\'e de Gen\`eve, D\'epartement de Physique Th\'eorique and CAP,
24 quai Ernest-Ansermet, CH-1211 Gen\`eve 4, Switzerland
}
\affiliation[c]{Coll\`ege de France, 11 Place M. Berthelot, 75005 Paris, France,\\
CERN, Theory Unit, Physics Department, CH-1211 Geneva 23, Switzerland}

\emailAdd{giuseppe.fanizza@ba.infn.it}
\emailAdd{gasperini@ba.infn.it}
\emailAdd{Giovanni.Marozzi@unige.ch}
\emailAdd{Gabriele.Veneziano@cern.ch}

\abstract{
We present a new method to compute the deflection of light rays in a perturbed FLRW geometry. We exploit the properties of the Geodesic Light Cone (GLC) gauge  where null rays propagate at constant angular coordinates irrespectively of the given (inhomogeneous and/or anisotropic) geometry. The gravitational deflection of null geodesics can then be obtained, in any other gauge, simply by expressing the angular coordinates of the given gauge in terms of the GLC angular coordinates. We apply this method to the standard Poisson gauge, including scalar perturbations, and give the full result for the deflection effect in terms of the 
direction of observation and observed redshift
up to second order, and up to third order for the leading lensing terms. We also compare our results with  those presently available in the  literature and, in particular, we provide a new non trivial check of a previous result on the luminosity-redshift relation up to second order in cosmological perturbation theory.
}

\keywords{cosmological perturbation theory, weak gravitational lensing, gravity 
\vskip13pt plus8pt minus11pt
\noindent{\bfseries\large\sffamily{Preprints:}} CERN-PH-TH-2015-132, BA-TH/696-15
}

\begin{document}

\maketitle


\section{Introduction}
\label{Sec1}
\setcounter{equation}{0}

It is well-known that the light-like signals emitted by the astrophysical sources propagate along the null geodesics of the cosmic geometry, and that the observed properties of this radiation (such as wavelength, polarization, propagation direction) are affected by the geometric properties of the large-scale space-time. For instance, if the geometry is expanding, then the frequency of the received radiation is shifted with respect to the emitted frequency according to the well known cosmological redshift effect.

Here we are interested, in particular, in the shift of direction of the received radiation with respect to the angular direction of the source, referred to a system of polar coordinates centered at the position of a static observer. More precisely, if we denote with $\theta^a_s$, $a=1,2$, the angular coordinates of the source in the given reference frame, and with $\theta^a_o$,  $a=1,2$, the polar angles controlling -- in the same frame -- the direction of the received radiation at the observer position, we are interested in computing the general relation $\theta^a_s= \theta^a_s (\theta^b_o)$, determined by the given model of cosmological geometry.

If the geometry is spatially homogeneous and isotropic the above relation, of course, is trivial: $\theta^a_s\equiv \theta^a_o$. However, if the cosmic geometry deviates, even perturbatively, from exact homogeneity and isotropy (for instance, because of macroscopic fluctuations of primordial inflationary origin, or because of the properties of the local matter distribution), then the light-like signals are geometrically deflected, and we have in general a non-trivial relation $\theta^a_s= \theta^a_s (\theta^b_o)$ dictated by the angular profile of the distorted shape of the null light-cone hypersurface.

The explicit form of such an angular relation depends not only on the given geometry but also on the chosen gauge. The aim of this paper is to present a new approach to the computation of the above angular relation in a generic (inhomogeneous/anisotropic) cosmological geometry, 
and apply this new method, in particular, to a perturbed FLRW metric which includes scalar perturbations up to third order parametrized in the usual Poisson gauge.

Our method is based on the coordinate transformation relating the angular coordinates of the source $\theta_s^a$, expressed in the Poisson gauge (PG), to the corresponding source coordinates $\tt_s^a$ expressed in the so-called geodesic light-cone (GLC) gauge
\cite{1}, namely on the transformation $\theta^a_s = \theta^a_s (\tt^b_s)$. Indeed, in  the GLC gauge,  null geodesics are characterized by constant values of the angular coordinates \cite{1,2}: it follows that 
$\tt^b_s\equiv \tt^b_o$, i.e. that the angular position of the source, $\tt^b_s$, always coincides in the GLC gauge with its ``apparent"  position  determined by the local direction of the light ray received by the observer. In addition, the coordinate transformation between PG and GLC gauge can always be defined in such a way that the angular coordinates of the two frames coincide (to all orders) at the observer position \cite{2,3}, namely at the origin of the PG system of polar coordinates: this implies $\tt_o^b \equiv \theta_o^b$. As a consequence, we have
\beq
\theta^a_s = GT (\tt^b_s)= GT ( \tt^b_o) \equiv GT (\theta^b_o )~,
\label{6}
\eeq
where $GT$ denotes the above-mentionad gauge transformation, with the dependence on $\tau$ not explicitly shown. Eq. (\ref{6})  provides the sought relation for the geometric deflection effect induced by the cosmic gravitational field, parametrized in the PG.

We recall that there is a large literature on the perturbative evaluation of cosmological observables, with and without the use of the GLC 
gauge.
Using the GLC coordinates, expressions for the luminosity distance-redshift relation were obtained in~\cite{3,Fanizza:2013doa,Marozzi:2014kua} (with applications discussed in~\cite{11a,BenDayan:2013gc,Ben-Dayan:2014swa}~\footnote{See \cite{Nugier:2013tca} for further details about the application of the GLC coordinates in the framework of the light-cone averaging procedure.}) and for the galaxy number counts in \cite{DiDio:2014lka}. 
Using different methods, similar expressions were obtained in~\cite{Umeh:2012pn,Umeh:2014ana} for the luminosity distance-redshift relation, and in \cite{Bertacca:2014dra,Bertacca:2014wga,Bertacca:2014hwa,Yoo:2014sfa} for the galaxy number counts.
Furthermore, results for the Newtonian density fluctuation were given in~\cite{Bernardeau:2001qr} and the study of lensing up to second order was discussed in~\cite{Bernardeau:2011tc}.

The paper is organized as follows. In Sect. \ref{Sec2} we briefly recall the main definitions and properties of the coordinate systems (GLC and PG) to be used in this paper. In Sect. \ref{Sec3} we first evaluate the cosmological deflection effect starting from the  transformation expressing GLC metric and coordinates in terms of the PG ones and then inverting the obtained transformation, up to the desired order. We shall use, to this purpose, the perturbative second-order results already presented in \cite{3}, 
expressing them in terms of observational coordinates and extending them to third perturbative order
(but only for  the leading lensing contributions, i.e. for those with the maximum number of angular derivatives).
In Sect. \ref{Sec4} we present our new method based on the  coordinate transformation directly expressing PG quantities in terms of GLC ones. Then, as an independent consistency check, we show that the results obtained for the deflection angle by this new method  exactly coincide with those of Sect. \ref{Sec3}. In Sect. \ref{Sec5} we show that our leading lensing terms satisfy, up to third order, a non-trivial ``lens equation" and compare our results with those currently available in the literature through different computational methods. 
 A few conclusive remarks are finally reported in Sect. \ref{Sec6}. 

In  Appendix \ref{AppA} we provide another application (and an additional test) of our new method by computing, up to the second perturbative order and with the approach of Sect. \ref{Sec4},  the full result  for the luminosity distance of the light source as a function of direction 
 of observation and redshift, and show that it exactly coincides with the one obtained in \cite{3,Fanizza:2013doa} using the approach of Sect. \ref{Sec3}. In Appendix \ref{AppB} we clarify the origin of an apparent disagreement on which averages of the luminosity distance are least affected by lensing. 
In Appendix \ref{AppC} we prove that, in spite of its form, the result of Appendix \ref{AppA} and of \cite{3,Fanizza:2013doa} for the luminosity distance-redshift relation is actually covariant.


\section{Geodesic Light-Cone and Poisson gauges: a short reminder}
\label{Sec2}
\setcounter{equation}{0}

In this Section we shortly  recall the main definitions and properties of the coordinate systems to be used in this paper.

The so-called geodesic light-cone (GLC) coordinates \cite{1}
are particularly adapted to describe signals that propagate along our past light-cone. They consist of a timelike  coordinate $\tau$,  of a null coordinate $w$, and of two angular coordinates $\tilde{\theta}^a$ ($a=1,2$). The parameter $\tau$ can  be  identified with synchronous gauge time \cite{2}, and thus provides the four-velocity of a  static geodesic observer in the form $u_{\mu} = -\partial_{\mu} \tau$. 
The line-element of the GLC metric reads:  
\beq
\label{LCmetric}
ds^2 =\Upsilon^2 dw^2 - 2 \Upsilon  dw d\tau+\gamma_{ab}(d \tilde{\theta}^a-U^a dw)(d \tilde{\theta}^b-U^b dw) ~~,~~~~~ a, b = 1,2~~,
 \eeq
or, in matrix form:
\beq
\label{GLCmetric}
g_{\mu\nu} =
\left(
\begin{array}{ccc}
0 & -\Upsilon &  \vec{0} \\
-\Upsilon & \Upsilon^2 + U^2 & -U_b \\
\vec0^{\,T}  &-U_a^T  & \gamma_{ab} \\
\end{array}
\right)
~~~~~,~~~~~
g^{\mu\nu} =
\left(
\begin{array}{ccc}
-1 & -\Upsilon^{-1} & -U^b/\Upsilon \\
-\Upsilon^{-1} & 0 & \vec{0} \\
-(U^a)^T/ \Upsilon & \vec{0}^{\, T} & \gamma^{ab}
\end{array}
\right) ~,
\eeq
where $\vec{0}= (0,0)$, $U_b=(U_1, U_2)$ and $U^2= \ga_{ab}U^aU^b$. Here 
$\Upsilon$,  $U^a$ and $\gamma_{ab} = \gamma_{ba}$ are arbitrary functions of the GLC coordinates, and $\gamma_{ab}$ and its inverse  $\gamma^{ab}$ lower and  raise  the two-dimensional indices. 

The $w=$ constant hypersurfaces define a foliation of space-time in terms of the null hypersurfaces ($\pa_\mu w \pa^\mu w=0$) that correspond to the past light-cones of a given observer throughout its history.
Also, in this gauge, the null geodesics connecting sources and observer are characterized  by the tangent vector $k^{\mu} = - \omega g^{\mu \nu} \partial_{\nu} w = -  \omega g^{\mu w} = \omega  \Upsilon^{-1} \delta^{\mu}_{\tau}$ (where $ \omega$ is an arbitrary normalization constant), meaning that photons  travel at constant values of $w$ and $\tilde{\theta}^a$. 
This crucial property of the GLC gauge will be extensively exploited in the following. It also renders the calculation of  the redshift  and luminosity distance (the two entries in the Hubble diagram) particularly simple.

Denoting by subscripts ``$o$'' and ``$s$'', respectively,  quantities evaluated at the observer and source space-time position, one finds that 
the exact  expression of  the redshift $z_s$ associated with a light ray going 
from ``$s$" to ``$o$" is  simply given (for  static observers and sources) by \cite{1}
\be
\label{redshift}
(1+z_s) = \frac{(k^{\mu} u_{\mu})_s }{(k^{\mu} u_{\mu})_o}  = \frac{(\partial^{\mu}w \pa_\mu \tau)_s }{(\partial^{\mu}w \pa_\mu \tau)_o}  = {\Ups(w_o, \tau_o, \ti \theta^a_o)\over \Ups(w_o, \tau_s, \ti \theta^a_s)} ~~.
\ee
Namely, it factorizes in terms of an entry of the GLC metric evaluated at the observer and the same quantity evaluated at the source, precisely as in the case of the FLRW metric. 
Similarly, an exact factorized expression for the so-called Jacobi Map~\cite{SEF}   was derived in \cite{Fanizza:2013doa}.
The determinant of the associated Jacobi matrix allows to express the luminosity (and area) distance, $d_L$ and $d_A$, entirely in
terms of $\gamma_{ab}$ and its derivatives at the observer, namely:
\beq
\label{lumdist}
d_L^2 =  (1+z_s)^4 d_A^2 = 
4  (1+z_s)^4 \frac{\sqrt{\gamma_s}}{\left[\det \left(u_{\tau}^{-1} \pa_\tau{\gamma}^{ab}\right) \gamma^{3/2}\right]_{o}} \,, 
\eeq
where $\gamma = \det\left(\gamma_{ab}\right)$. The above result has been used 
to obtain non-perturbative expressions for weak lensing quantities
such as magnification, convergence, shear and vorticity \cite{Fanizza:2014baa}.

In order to connect our results to those usually presented in the literature we  also conveniently introduce the so-called Poisson gauge (PG)~\cite{PG} (sometimes referred to, at first order,  as the ``Newtonian" or ``longitudinal" gauge).
Neglecting  vector and tensor perturbations, the PG metric~takes the  form:
\beq
ds^2= a^2(\eta) \Big[ -  d \eta^2  \left(1+2 \Phi\right)+ \left(1-2 \Psi \right) \left(dr^2 +r^2 d\theta^2 + r^2 \sin^2 \theta d\phi^2\right)\Big].
\label{1}
\eeq
Here $a(\eta)$ is the scale factor, $\eta$ is the conformal time, and the (generalized) Bardeen potentials $\Phi(\eta, r, \theta^a)$ and $\Psi(\eta, r, \theta^a)$ describe the first-order ($\varphi$, $\psi$),   second-order ($\varphi^{(2)}$, $\psi^{(2)}$) and third order ($\varphi^{(3)}$, $\psi^{(3)}$) scalar perturbations of a conformally flat FLRW metric background:
\beq
\Phi= 
\varphi +{1\over 2} \varphi^{(2)}+\frac{1}{6} \varphi^{(3)}, ~~~~~~~~~~~~~ \Psi=\psi +{1\over 2} \psi^{(2)}+\frac{1}{6} \psi^{(3)}.
\label{2}
\eeq
Assuming the absence of sources with anisotropic stresses we shall  put everywhere $\varphi= \psi$, so that, to first order, $\Phi= \Psi= \psi$. 

For the application discussed in Sec. IV it is also useful to redefine the Poisson Gauge using the light-cone-like (but not exactly null) coordinate $\ep= \eta+r$, so that the PG metric (\ref{1}) takes the alternative form:
\bea
ds^2 = g_{\mu\nu}^{\rm PG} dy^\mu d y^\nu &=& a^2(\eta) \Big[ -  2d \eta^2 \left(\Phi+\Psi \right) + \left(1-2 \Psi\right) \left(d{\ep}^2 - 2 d\eta d \ep \right) 
\nonumber \\
& & 
+ \left(1-2 \Psi\right) \left(\ep-\eta\right)^2 \left( d\theta^2 + \sin^2 \theta d \phi^2 \right)\Big].
\label{3}
\eea
The inversion of Eq. (\ref{3}), including the contribution of scalar perturbations up to third order, then gives:
\beq
g_{\rm PG}^{\mu\nu} ={1\over a^2}
\left(
\begin{array}{ccc}
-1 +A & -1+A & \vec{0} \\
-1+A & B& \vec{0} \\
\vec{0}^{\, T} & \vec{0}^{\, T} & \gamma^{ab}_{\rm PG}
\end{array}
\right) ~~~,
\label{8}
\eeq
where
\bea
&&
A= 2 \psi - 4 \psi^2 + \vp^{(2)} + \frac{1}{3} \vp^{(3)}-4 \psi \vp^{(2)} + 8 \psi^3, 
\nonumber \\ &&
 B= 4 \psi + \psi^{(2)}+ \vp^{(2)}
+\frac{1}{3} \left(\vp^{(3)}+\psi^{(3)}\right) -4 \psi  \vp^{(2)}+16 \psi^3 +4 \psi \psi^{(2)},
\nonumber \\ &&
\gamma^{ab}_{\rm PG}= \left(1+2 \psi +4 \psi^2 +\psi^{(2)}
+8 \psi^3 +4 \psi \psi^{(2)}+\frac{1}{3} \psi^{(3)}
\right) \ga_0^{ab}, 
\nonumber \\ &&
 \ga_0^{ab}= \left(\ep-\eta \right)^{-2} {\rm diag} \left(1, \sin^{-2} \theta \right),
\label{9}
\eea
and where the scalar fluctuations  $\psi, \psi^{(2)}, \vp^{(2)}, \psi^{(3)}, \vp^{(3)}$  are functions of the PG coordinates $\eta, \ep, \theta^a$. 

We are now ready to carry out the connection between coordinates and metric in these different frames, and to compute the deflection angles in the PG. In the next two Sections we shall adopt two quite different procedures: the first  is the one followed in our previous papers \cite{1,2,3,Fanizza:2013doa,Marozzi:2014kua}; the second   is conceptually (and also, to some extent, technically) simpler. In spite of being very different the two procedures will be shown to give exactly equivalent results.


\section{Expressing GLC quantities in terms of PG coordinates and physical observables}
\label{Sec3}
\setcounter{equation}{0}
In order to evaluate the light deflection induced by scalar cosmological perturbations we will consider, in this section, the transformation relating GLG coordinates and metric to those of the Poisson gauge, and  then  express the result in terms of the observer's detection angle and redshift. We shall use the approach  already investigated to second order in \cite{3,Fanizza:2013doa} (for vanishing anisotropic stress) and in \cite{Marozzi:2014kua} (for non-vanishing anisotropic stress), extending the previous computations 
in order to describe the light deflection in terms of observational variables and to include the leading lensing contributions up to third order.

 \subsection{The coordinate transformation}
\label{Sec31}

Let us consider the general (exact) transformation connecting the (inverse of the) GLC metric (\ref{GLCmetric}) to the (inverse of the) PG metric (\ref{1}),
\be
g_{GLC}^{\rho\sigma}(x)=\frac{\partial x^\rho}{\partial y^\mu}
\frac{\partial x^\sigma}{\partial y^\nu} g_{PG}^{\mu\nu}(y) ~~,
\label{EqBetweenGauges}
\ee
where we have denoted by $y^\mu=(\eta, r, \theta^a)$ the PG coordinates and by $x^\nu=(\tau, w,\ti{\theta}^a)$ the GLC ones. Solving this set of differential equations for the variables $\tau, w, \tt^a$ (with the boundary conditions that the transformation is non singular, and that $\tt^a= \theta^a$ at the observer position $r=0, \eta=\eta_o$), we obtain \cite{3,Fanizza:2013doa}, up to second order in the scalar perturbations:
\beq
\tau = \tau^{(0)}+\tau^{(1)}+\tau^{(2)},
\eeq 
with
\bea
&&
\tau^{(0)} = \left( \int_{\eta_{in}}^\eta d\eta' a(\eta') \right), ~~~~~~~~~~~~~ \tau^{(1)} = a(\eta) P(\eta, r, \theta^a)\,, \nonumber \\
& & \tau^{(2)}= \int_{\eta_{in}}^\eta d\eta' \frac{a(\eta')}{2} \left[ \phi^{(2)} - \psi^2 + ( \partial_r P )^2 + \gamma_0^{ab} ~ \partial_a P ~ \partial_b P \right] (\eta', r, \theta^a)~;
\label{tau2order} 
\eea
\beq
w = w^{(0)}+w^{(1)}+w^{(2)},
\eeq
with
\bea
&&
 w^{(0)}=\eta_+ , ~~~~~~~~~~~~~~~~~~~~~ w^{(1)}=Q(\eta_+, \eta_-, \theta^a) \,,
\nonumber \\
& &  
w^{(2)}= { \frac{1}{4} \int_{\eta_o}^{\eta_-} dx~ \left[ {\psi}^{(2)} + {\phi}^{(2)} + 4 \psi  \partial_+ Q + {\gamma}_0^{ab} ~ \partial_a Q ~ \partial_b Q \right] (\eta_+, x, \theta^a)}\,;
\label{w2order} 
\eea
\beq
\tilde{\theta}^a = \tilde{\theta}^{a (0)}+\tilde{\theta}^{a (1)}+\tilde{\theta}^{a (2)} ,
\eeq
with 
\bea
&&
 \tilde{\theta}^{a (0)}=\theta^a , ~~~~~~~~~~~~~~~~~~~~~
\tilde{\theta}^{a (1)}={\frac12 \int_{\eta_o}^{\eta_-} dx~ \left[ {\gamma}_0^{ab} \partial_b Q \right] (\eta_+,x,\theta^a)} 
\nonumber \\
& & 
\tilde{\theta}^{a (2)}= { \int_{\eta_o}^{\eta_-} dx~ 
\left[ 
\frac{1}{2}
{\gamma}_0^{ac} \partial_c w^{(2)} 
+ {\psi} {\gamma}_0^{ac} \partial_c w^{(1)} 
+\frac{1}{2}{\gamma}_0^{dc} \partial_c w^{(1)} \partial_d   \tilde{\theta}^{a (1)}
+\psi\partial_+\tilde{\theta}^{a (1)}
\right. } \nonumber \\
& & \left.~~~~~~~~~~~~~~~~~~~~~~~~~~~  -\partial_+ w^{(1)} \partial_-\tilde{\theta}^{a (1)} \right](\eta_+,x,\theta^a) \,.
\label{thetatilde2orderShort}
\eea
We have defined $\ga_0^{ab} = {\rm diag}(r^{-2},r^{-2} \sin^{-2}\theta)$, and we recall that the lower integration limit $\eta_{in}$ represents an early enough time when the perturbations (or better their integrands) were negligible.
We have also used the zeroth-order light-cone variables $\eta_\pm= \eta \pm r$, 
with corresponding partial derivatives:
\beq
\pa_\eta = \pa_+ + \pa_- ~~~,~~~~~ \pa_r = \pa_+ - \pa_- ~~~,~~~~~\pa_\pm= {\pa \over \pa \eta_\pm}={1\over 2} \left( \pa_\eta \pm \pa_r \right) ~~.
\label{partial+-}
\eeq
Finally, we have defined
\be
P(\eta, r, \theta^a) = \int_{\eta_{in}}^\eta d\eta' \frac{a(\eta')}{a(\eta)} \psi(\eta',r,\theta^a)
\,\,\,\,,\,\,\,\, ~~~ Q(\eta_+, \eta_-, \theta^a) = \int_{\eta_o}^{\eta_-} dx~  \psi(\eta_+,x,\theta^a) ~.
\label{PQ}
\ee

Using the above results we can then compute the non-trivial entries of the (inverse)  GLC metric and we find, up to second order  \cite{3,Fanizza:2013doa}:
\bea
\Upsilon^{-1} &=& \frac{1}{a(\eta)} \left[ 1 + \partial_+ Q - \partial_r P  
+\partial_{\eta}w^{(2)} + \frac{1}{a}(\partial_\eta - \partial_r) \tau^{(2)} - \phi^{(2)} + \psi^2 
\right. \nonumber \\
& & \left.
-\psi \left( \partial_+ Q+\partial_r P\right) - \partial_r P\partial_+ Q
- \gamma^{ab}_0 \partial_a P \partial_b Q\right] \,,
\label{Ups1}  
\\
U^a &=& \partial_{\eta}\tilde{\theta}^{a (1)}-\frac{1}{a}\gamma_0^{ab}\partial_b \tau^{(1)}+\partial_{\eta}\tilde{\theta}^{a (2)}-
\frac{1}{a}\gamma^{ab}_0\partial_b \tau^{(2)} - \frac{1}{a} \partial_r \tau^{(1)} \partial_r \tilde{\theta}^{a(1)}  \cr
& &  -\psi \partial_\eta \tilde{\theta}^{a (1)}-\frac{2}{a} \psi \gamma^{ab}_0 \partial_b \tau^{(1)} 
-\frac{1}{a}\gamma_0^{cd}\partial_c \tau^{(1)} \partial_d \tilde{\theta}^{a (1)} \nonumber \\
& &
+\left(\partial_+ Q - \partial_r P\right) \left(-\partial_{\eta} \tilde{\theta}^{a (1)}+\frac{1}{a}\gamma^{ab}_0 \partial_b \tau^{(1)}\right)\,,
\label{Ua1}
\\ 
\gamma^{ab} &=& a^{-2}\left\{ \gamma_0^{ab} \left(1 +  2 \psi\right) +\left[\gamma_0^{a c} \partial_c \tilde{\theta}^{b (1)}+ (a\leftrightarrow b) \right]+ \gamma_0^{ab}
\left(\psi^{(2)} + 4 \psi^2 \right)-\partial_\eta \tilde{\theta}^{a (1)}\partial_\eta \tilde{\theta}^{b (1)}
\right. \nonumber \\
& & \left.
+\partial_r \tilde{\theta}^{a (1)}\partial_r \tilde{\theta}^{b (1)} +2 \psi \left[\gamma_0^{a c} \partial_c \tilde{\theta}^{b (1)}+ (a\leftrightarrow b) \right]+\gamma_0^{c d} \partial_c \tilde{\theta}^{a (1)}
 \partial_d \tilde{\theta}^{b (1)} \right.
 \nonumber \\
& &  \left.
 +\left[\gamma_0^{a c} \partial_c \tilde{\theta}^{b (2)}+ (a\leftrightarrow b) \right] \right\}.
\label{gammaab}
\eea 


\subsection{Light deflection to second order}
\label{Sec32}

As already stressed in the Introduction, we need now to express the angular position of the source, given by the PG coordinates $\theta_s^a$, as a function of the corresponding GLG coordinates $\tt_s^a$ which exactly identify the ``observed" position of the source, $\theta_o^a$, determined by the direction of the received radiation as measured by a static PG observer (see Eq. (\ref{6})). We also want to give the final result for $\theta_s^a$ in terms of observational variables only: the above mentioned angular direction $\theta_o^a$, and the measured redshift of the source, $z_s$.

In order to complete the first step one has to invert the coordinate transformation given in Eq. (\ref{thetatilde2orderShort}). This was done, for the first time and up to second order, in \cite{3}. The result we obtained,
\bea
{\theta}^a &=& {\theta}^{a (0)}+{\theta}^{a (1)}+{\theta}^{a (2)}
= \tilde{\theta}^a - \frac12 \int_{\eta_o}^{\eta_-} dx~ \gamma_0^{ab} \int_{\eta_o}^x dy~ \partial_b \psi(\eta_+,y,\tilde{\theta}^a)\nonumber \\
& &
+\frac{1}{4}\left[\int_{\eta_o}^{\eta_-} dx~ \gamma_0^{cb} \int_{\eta_o}^x dy~ \partial_b \psi(\eta_+,y,\tilde{\theta}^a)\right]
\partial_c \left[\int_{\eta_o}^{\eta_-} dx~ \gamma_0^{ad} \int_{\eta_o}^x dy~ \partial_d \psi(\eta_+,y,\tilde{\theta}^a)\right]
\nonumber \\
& &-\int_{\eta_o}^{\eta_-} dx~ \left[ \gamma_0^{ac} \zeta_c + \psi ~ \xi^a + \lambda^a \right] (\eta_+,x,\tilde{\theta}^a) ~~
\label{thetatilde2orderShort_Inverted}
\eea
(see Eq. (3.30) of \cite{3}), can be conveniently expressed in conformal time $\eta$, and written for the angular coordinate of the source as ${\theta}^a_s = {\theta}^{a (0)}_s+{\theta}^{a (1)}_s+{\theta}^{a (2)}_s$, 
with ${\theta}^{a (0)}_s= \tt_s^a \equiv \theta_o^a$, and with 
\bea
\theta^{a (1)}_s&=& -2 \int_{\eta_s}^{\eta_o}d\eta' \ga_0^{ab} \int_{\eta'}^{\eta_o}d\eta'' \pa_b  \psi (\eta'') \, ,
\label{thetatilde1order_Inverted}
\\
\theta^{a (2)}_s&=&4\,\pa_b\left[ \int_{\eta_s}^{\eta_o}d\eta'\ga_0^{ac}\pa_c\int_{\eta'}^{\eta_o}d\eta''\psi(\eta'') \right] \int_{\eta_s}^{\eta_o}d\eta'\ga_0^{bd}\pa_d\int_{\eta'}^{\eta_o} d\eta''\psi(\eta'')\nonumber\\
& & +2\int_{\eta_s}^{\eta_o}d\eta'\left[ \ga_0^{ac}\zeta_c(\eta')+\psi(\eta')\xi^a(\eta')+\lambda^a(\eta') \right],
\label{thetatilde2order_Inverted}
\eea
where
\bea
\zeta_c(\eta')&=&-\frac{1}{4}\int_{\eta'}^{\eta_o}d\eta''\pa_c\left\{ \psi^{(2)}+\phi^{(2)}+4\,\psi\left[ \psi_o-\psi-2\int_{\eta''}^{\eta_o}d\eta'''\pa_{\eta'''}\psi(\eta''') \right]\right.
\nonumber\\
&&\left. +4 \ga_0^{ab}\pa_a\left(\int_{\eta''}^{\eta_o}d\eta'''\psi(\eta''')\right)\,\pa_b\left(\int_{\eta''}^{\eta_o}d\eta'''\psi(\eta''')\right) \right\},
\label{zeta1}
\\
\xi^a(\eta')&=&-\lim_{\eta\rightarrow \eta_o}\left( \ga_0^{ab}(\eta)\pa_b\int_{\eta}^{\eta_o}d\eta''\psi(\eta'') \right)-\ga_0^{ab}(\eta')\pa_b\int_{\eta'}^{\eta_o}d\eta''\psi(\eta'')\nonumber\\
&& -\int_{\eta'}^{\eta_o}d\eta''\ga_0^{ab}\pa_b\left[ \psi_o-2\int_{\eta''}^{\eta_o}d\eta'''\pa_{\eta'''}\psi(\eta''') \right],
\label{xi2}\\
\lambda^a(\eta')&=&-2\ga_0^{dc}\pa_c\left( \int_{\eta'}^{\eta_o}d\eta''\psi(\eta'') \right)\int_{\eta'}^{\eta_o}d\eta''\pa_d\left[ \ga_0^{ab}\pa_b\left( \int_{\eta''}^{\eta_o}d\eta'''\psi(\eta''') \right) \right]\nonumber\\
&&+\left[ \psi_o-\psi(\eta')-2\int_{\eta'}^{\eta_o}d\eta''\pa_{\eta''}\psi(\eta'') \right]\ga_0^{ab}\pa_b\int_{\eta'}^{\eta_o}d\eta''\psi(\eta'')\,.
\label{lambda3}
\eea
Note that, in the above equations and hereafter, partial derivatives like 
$\partial_a$ are made with respect to $\tilde{\theta}^a=\theta_o^a$. Also, for brevity, we only indicate the
conformal time $\eta$ as argument on the  integrands over the line-of-sight, suppressing other arguments like $ \tilde{\theta}^a$.

According to  Eq. (\ref{thetatilde1order_Inverted}), we can also introduce a so-called (first order) lensing potential (see for instance \cite{Lewis:2006fu}), defined by 
\be
\psiP(\eta)=-\frac{2}{\eta_o-\eta} \int_{\eta}^{\eta_o} d\eta' \frac{\eta'-\eta}{\eta_o-\eta'} \psi(\eta')=  
-\frac{2}{r} \int_0^r dr' \frac{r-r'}{r'} \psi(r')
\label{Lensing-Potential}
\ee
and such that, to first order,  $\theta_s^a=\hat{\gamma}_0^{ab} \partial_b \psiP(\eta_s)$, where $\hat{\gamma}_0^{ab} =r^2 \gamma_0^{ab}$.

In view of possible phenomenological applications it may be appropriate, at this point, to evaluate the terms providing the leading lensing contributions at large enough redshift ($z \gaq 0.3$),  to second order in the deflection angle. 
Let us note, to this purpose, that the corrections arising from writing the deflection angles in terms of the observed redshift (see below)
are subleading, hence they will be neglected.
Considering Eqs. (\ref{thetatilde2order_Inverted})-(\ref{lambda3}), let us first select those terms contributing to $\theta^{a (2)}_s$ and containing the maximum number of transverse derivatives  (namely, {\em three} angular derivatives). Denoting such terms with $\theta^{a (2)}_{s,A}$ we have:
\bea
\theta^{a (2)}_{s,A}&=&
4 \left[\int_{\eta_s}^{\eta_o}d\eta'\,\ga_0^{cd}\pa_d\int_{\eta'}^{\eta_o}d\eta''\psi(\eta'')\right]\,
\partial_c \int_{\eta_s}^{\eta_o}d\eta'\,\ga_0^{ab}\pa_b\int_{\eta'}^{\eta_o}d\eta''\psi(\eta'')
\nonumber \\ & &
- 4\int_{\eta_s}^{\eta_o}dx\left[ \ga_0^{bc}\,\pa_b\left(\int_{\eta'}^{\eta_o}d\eta''\psi(\eta'')\right)\,\pa_c\left( \int_{\eta'}^{\eta_o}d\eta''\,\ga_0^{ab}\pa_b\int_{\eta''}^{\eta_o}d\eta'''\psi(\eta''') \right)\right.\nonumber\\
&&\left.+\frac{1}{2}\ga_0^{ab}\,\pa_b\int_{\eta'}^{\eta_o}d\eta''\,\ga_0^{cd}\,\pa_c\left(\int_{\eta''}^{\eta_o}d\eta'''\,\psi(\eta''')\right)\,\pa_d\left(\int_{\eta''}^{\eta_o}d\eta'''\,\psi(\eta''')\right) \right]\,.
\label{theatas-S-A-form1}
\eea 
After some algebraic manipulations and integrations by parts one finds that the above equation can be rewritten  in a somewhat more useful form as:
\bea
&&
\theta^{a (2)}_{s,A}=-\frac{2}{\eta_o-\eta_s}\int_{\eta_s}^{\eta_o}d\eta'\frac{\eta'-\eta_s}{\eta_o-\eta'}\hat\ga_0^{ac}\pa_c\pa_b\psi(\eta')\,\hat\ga_0^{bd}\pa_d{\psiP}(\eta')\nonumber\\
&+&\frac{2}{\eta_o-\eta_s}\int_{\eta_s}^{\eta_o}d\eta'\frac{\eta'-\eta_s}{\eta_o-\eta'}\hat\ga_0^{dc}\pa_c\psi(\eta')\left( \pa_d\hat\ga_0^{ab}\right)\pa_b{\psiP}(\eta')\nonumber\\
&+&\int_{\eta_s}^{\eta_o}\frac{d\eta'}{\left(\eta_o-\eta'\right)^2}\int_{\eta'}^{\eta_o}\frac{d\eta''}{\left( \eta_o-\eta'' \right)^2}\left[  -2\hat\ga_0^{ac}\left( \pa_c\hat\ga_0^{db} \right)\pa_d\left(\g{\eta''}{\eta'''}\right)\pa_b\left(\g{\eta''}{\eta'''}\right)\right.\nonumber\\
&-&\left.4\hat\ga_0^{dc}\left( \pa_d\hat\ga_0^{ab} \right)\pa_c\left(\g{\eta''}{\eta'''}\right)\pa_b\left(\g{\eta''}{\eta'''}\right)\right]\nonumber\\
&+&\left( \pa_b\hat\ga_0^{ac} \right)\hat\ga_0^{bd}\pa_c{\psiP}(\eta_s)\pa_d{\psiP}(\eta_s)
\label{theatas-S-A-form1-simplified}
\eea
Finally, neglecting  the subleading terms with only two angular derivatives acting on the metric perturbations, one obtains for the second-order leading  contributions  the following simple expression:
\bea
(\theta^{a (2)}_{s})_{\rm Leading}&=&-\frac{2}{\eta_o-\eta_s} \int_{\eta_s}^{\eta_o}d\eta' \frac{\eta'-\eta_s}{\eta_o-\eta'} \hat{\gamma}_0^{a b} 
\hat{\gamma}_0^{c d} \partial_d \psiP(\eta')
\partial_c \partial_b \psi(\eta')
\nonumber \\
&=&-\frac{2}{\eta_o-\eta_s} \int_{\eta_s}^{\eta_o}d\eta' \frac{\eta'-\eta_s}{\eta_o-\eta'} \hat{\gamma}_0^{a b} \theta^{c (1)}(\eta')
\partial_c \partial_b \psi(\eta')\,.
\label{LeadStorder2}
\eea
This is in perfect agreement with the recent result obtained in \cite{Bonvin:2015uha}.

Let us now express the full second-order result for $\theta_s^a$ in terms of the redshift $z_s$. Following the procedure introduced in previous papers 
 \cite{2,3,Marozzi:2014kua} we consider the convenient expansion $\eta_s=\bar{\eta}_s^{(0)}+\bar{\eta}_s^{(1)}+ \cdots$ and $r_s=\bar{r}_s^{(0)}+\bar{r}_s^{(1)}+ \cdots$, where the zeroth-order parameters satisfy the condition $\bar{\eta}_s^{(0)}+\bar{r}_s^{(0)}=w_o=\eta_o$ and are defined in such a way that the observed redshift $z_s$ is given by
\beq
1+z_s=\frac{a(\eta_o)}{a(\bar{\eta}_s^{(0)})} .
\label{red}
\eeq
By expanding around $\bar{\eta}_s^{(0)}$, $\bar{r}_s^{(0)}$ the exact expression  (\ref{redshift})  for $z_s$, using Eq. (\ref{Ups1}) for $\Ups^{-1}$ and the definition (\ref{red}) for $z_s$, and expanding to first order also the light-cone condition $w=w_o=$ const, one  finds that the first-order corrections $\bar{\eta}_s^{(1)}$, $\bar{r}_s^{(1)}$ are given by \cite{3,Marozzi:2014kua}:
\bea
&&
\bar{\eta}_s^{(1)} = {1\over \mathcal{H}_s}  \left[
 v_{||o}- v_{||s}+\psi_o-\psi_s -2 \int_{\bar{\eta}_s^{(0)}} ^{\eta_o} d \eta' \partial_{\eta'}\psi\left(\eta'\right) \right],
\label{Redshift1}\\
&&
\bar{r}_s^{(1)}=-\bar{\eta}_s^{(1)}+2 
 \int_{\bar{\eta}_s^{(0)}}^{\eta_o} d \eta' \psi\left(\eta'\right)\,,
 \eea
where we have defined $\mathcal{H}_s=(\pa_\eta a /a )_{\bar{\eta}_s^{(0)}}$ and we recall that $v_{||s} \equiv (\pa_r P) (\bar{\eta}_s^{(0)}, \bar{r}_s^{(0)}, \theta)$, with $P$ given by Eq. (\ref{PQ}), is the first-order velocity perturbation projected along the radial direction connecting source and observer (see also \cite{Fanizza:2013doa}).

The desired expression for $\theta_s^a$ in terms of the observation variables $\theta_o^a$ and $z_s$ can now be obtained by Taylor expanding the coordinate transformation (\ref{thetatilde2orderShort_Inverted}), up to second order, as follows:
\bea
\theta_s^a (z_s, \theta_o) &=& \theta_o^a + \theta^{a (1)} (\bar{\eta}_s^{(0)}, \bar{r}_s^{(0)}, \theta_o)+  \theta^{a (2)}  (\bar{\eta}_s^{(0)}, \bar{r}_s^{(0)}, \theta_o)
\nonumber \\
&+&\bar{\eta}_s^{(1)} (\pa_\eta \theta^a)^{(1)}   (\bar{\eta}_s^{(0)}, \bar{r}_s^{(0)}, \theta_o)
+\bar{r}_s^{(1)} (\pa_r \theta^a)^{(1)}   (\bar{\eta}_s^{(0)}, \bar{r}_s^{(0)}, \theta_o)+ \cdots.
\label{Taylor}
\eea
Here $ \theta^{a (1)}$ and $ \theta^{a (2)}$ are given, respectively, by
Eqs. (\ref{thetatilde1order_Inverted}) and (\ref{thetatilde2order_Inverted}) evaluated at $\bar{\eta}_s^{(0)}, \bar{r}_s^{(0)}, \theta_o^a$. The time parameter $\bar{\eta}_s^{(0)}$, in  turn, is directly related  to the observed redshift by Eq. (\ref{red}), namely by the standard relation $dz_s= - (1+z_s) \mathcal{H}_s d \bar{\eta}_s^{(0)}$. 

The first order part of Eq. (\ref{Taylor}) then gives:
\beq
\theta^{a (1)}_s (z_s, \theta_o)= -2 \int_{\bar{\eta}^{(0)}_s}^{\eta_o}d\eta' \ga_0^{ab} \int_{\eta'}^{\eta_o}d\eta'' \pa_b  \psi (\eta'') 
\label{FinalThetaStorder1}
\eeq
(see also Eq. (\ref{thetatilde1order_Inverted})). 
Considering the second-order part of Eq. (\ref{Taylor}), and keeping all (leading as well as non leading) terms, we  obtain:
\bea
\left[\theta^{a (2)}_s (z_s, \theta_o) \right]_{\rm Full}
&=&4\,\pa_b\left[ \int_{\bar{\eta}_s^{(0)}}^{\eta_o}d\eta'\ga_0^{ac}\pa_c\int_{\eta'}^{\eta_o}d\eta''\psi(\eta'') \right] \int_{\bar{\eta}_s^{(0)}}^{\eta_o}d\eta'\ga_0^{bd}\pa_d\int_{\eta'}^{\eta_o} d\eta''\psi(\eta'')\nonumber\\
& &+2\left\{ \frac{1}{\mathcal{H}_s}\left[ v_{o\rVert}-v_{s\rVert}+\psi_o-\psi_s-2\int_{\bar{\eta}_s^{(0)}}^{\eta_o}d\eta'\pa_{\eta'}\psi(\eta') \right]
\right.  
\nonumber \\ 
& &  \left.
-\int_{\bar{\eta}_s^{(0)}}^{\eta_o}d\eta'\psi(\eta')\right\}\frac{1}{\left( \eta_o-\eta_s \right)^2}\int_{\bar{\eta}_s^{(0)}}^{\eta_o}d\eta'\,\hat\ga_0^{ab}\pa_b\psi(\eta')\nonumber\\
& &+4\int_{\bar{\eta}_s^{(0)}}^{\eta_o}d\eta'\psi(\eta')\int_{\bar{\eta}_s^{(0)}}^{\eta_o}d\eta'\frac{1}{\left( \eta_o-\eta' \right)^3}\int_{\eta'}^{\eta_o}d\eta''\hat\ga_0^{ab}\pa_b\psi(\eta'')\nonumber\\
& &-4\int_{\bar{\eta}_s^{(0)}}^{\eta_o}d\eta'\,\psi(\eta')\int_{\bar{\eta}_s^{(0)}}^{\eta_o}d\eta'\ga_0^{ab}\pa_b\left[ \frac{\psi(\eta')-\psi_o}{2}+\int_{\eta'}^{\eta_o}d\eta''\,\pa_{\eta''}\psi(\eta'') \right]\nonumber\\
& &+2\int_{\bar{\eta}_s^{(0)}}^{\eta_o}d\eta'\left[ \ga_0^{ac}\zeta_c(\eta')+\psi(\eta')\xi^a(\eta')+\lambda^a(\eta') \right]\,
\label{FinalThetaStorder2}
\eea
where $\zeta, \xi, \la$ are given by Eqs.(\ref{zeta1})--(\ref{lambda3}). 
Hereafter we shall omit the terms from the velocity and the gravitational potential at the observer position,
i.e. $v_{||o}$ and $\psi_o$, since, in any case, 
the first can be removed by going to the CMB frame and, in general, they cannot  be evaluated within perturbation theory.

After some manipulations and integration by parts we can finally rewrite our complete second-order result in the following simpler form
\bea
&&
\left[\theta^{a (2)}_s (z_s, \theta_o) \right]_{\rm Full}=
\theta^{a (2)}_{s,A} (\bar{\eta}_s^{(0)}, \bar{r}_s^{(0)}, \theta_o)
\nonumber\\ &&
-{2\over\mathcal{H}_s} \left[ v_{s\rVert} + \psi_s+2\int_{\bar{\eta}_s^{(0)}}^{\eta_o}d\eta'\pa_{\eta'}\psi(\eta') \right]\ga_0^{ab}\pa_b\g{\bar{\eta}_s^{(0)}}{\eta'}\nonumber\\
& &-4\left(\g{\bar{\eta}_s^{(0)}}{\eta'}\right)\ga_0^{ab}\pa_b\g{\bar{\eta}_s^{(0)}}{\eta'}\nonumber\\
& &-\frac{1}{2}\int_{\bar{\eta}_s^{(0)}}^{\eta_o}d\eta'\ga_0^{ab}\pa_b\int_{\eta'}^{\eta_o}d\eta''\left\{ \psi^{(2)}+\phi^{(2)}-4\psi(\eta'')\left[\G{\eta''}{\eta'''}\right] \right\}\nonumber\\
& &+2\int_{\bar{\eta}_s^{(0)}}^{\eta_o}d\eta'\ga_0^{ab}\pa_b\psi(\eta')\g{\eta'}{\eta''}
-2\g{\bar{\eta}_s^{(0)}}{\eta'}\ga_0^{ab}\pa_b\g{\eta'}{\eta''}\nonumber\\
& &-2\int_{\bar{\eta}_s^{(0)}}^{\eta_o}d\eta'\ga_0^{ab}\pa_b\left[ \left(\G{\eta'}{\eta''}\right)\g{\eta'}{\eta''} \right],
\label{Final-theta2-general}
\eea
where $\theta^{a (2)}_{s,A}$ is given by Eq.(\ref{theatas-S-A-form1-simplified}) evaluated for $\eta_s= \bar{\eta}_s^{(0)}$ and $r_s= \bar{r}_s^{(0)}$.


\subsection{Leading third-order contributions}
\label{3.4}

The above computation will now be extended to describe the angular deflection of light-like signals up to the third perturbative order. We are interested, in particular, in the leading lensing contribution corresponding, at third order, to angular corrections generated by {\em five} angular derivatives of the perturbed cosmological metric.

It will be enough to that purpose to consider only the leading terms of the coordinate transformation introduced in Sect. \ref{Sec31} for $w$ and $\tt^a$. Let us recall here, for convenience, that such terms are given by
\bea
w^{(0)}&=&\eta^+ ~,~  w^{(1)}=Q(\eta_+,\eta_-,\theta^a), \nonumber\\
w^{(2)}&=&\frac{1}{4}\int_{\eta_o}^{\eta_-}dx\,\left[\ga_0^{ab}\,\pa_aQ\,\pa_bQ\right](\eta_+,x,\theta^a),
\eea
and by 
\bea
\tilde{\theta}^{a (0)}&=&\theta^a ~,~
\tilde{\theta}^{a (1)} = \frac{1}{2}\int_{\eta_o}^{\eta_-}dx\,\left[\ga_0^{ab}\pa_bQ\right](\eta_+,x,\theta^a),
\label{Theta1-lead}\\
\tilde{\theta}^{a (2)} &=&\frac{1}{2}\int_{\eta_o}^{\eta_-}dx\left[ \ga_0^{bc}\,\pa_bQ\,\pa_c
\tilde{\theta}^{a (1)}+\ga_0^{ab}\,\pa_bw^{(2)} \right](\eta_+,y,\theta^a).
\label{Theta2-lead}
\eea
Extending to third order the above results  we obtain the following leading angular contributions:
\bea
\tilde{\theta}^{a (3)}&=&\frac{1}{2}\int_{\eta_o}^{\eta_-}dx\left[ \ga_0^{bc}\left(\pa_bw^{(2)}\pa_c
\tilde{\theta}^{a (1)}+\pa_bQ\,\pa_c
\tilde{\theta}^{a (2)}\right)+\ga_0^{ab}\,\pa_bw^{(3)} \right](\eta_+,x,\theta^a),
\label{3.33}
\eea
where
\beq
w^{(3)}=\frac{1}{8}\int_{\eta_o}^{\eta_-}dx\,\left[\ga_0^{ab}\,\pa_aQ\right](\eta_+,x,\theta^a)\,\pa_b\int_{\eta_o}^xdy\,\left[\ga_0^{cd}\pa_cQ\pa_dQ\right](\eta_+,y,\theta^a)\,.
\label{w(3)}
\eeq

As before, we need  to invert the above transformation in order to obtain $\theta^a_s$ as a function of the observation angle $\tt_s^a \equiv \theta_o^a$. However, as already mentioned, there is no need of performing also the redshift expansion (as done in the previous section) since  that expansion would only produce subleading contributions. 

By inverting and Taylor expanding Eq. (\ref{3.33}), order by order, around 
$\theta_o^a$ we then obtain, for the angular coordinate of the source,
\be
(\theta^{a (3)}_{s})_{\rm Leading}=\tilde{\theta}^{b (2)}_s \partial_b \tilde{\theta}^{a (1)}_s-\tilde{\theta}^{c (1)}_s \partial_c \tilde{\theta}^{b (1)}_s 
\partial_b \tilde{\theta}^{a (1)}_s-\frac{1}{2} \tilde{\theta}^{b (1)}_s  \tilde{\theta}^{c (1)}_s \partial_b \partial_c  \tilde{\theta}^{a (1)}_s
+ \tilde{\theta}^{b (1)}_s  \partial_b \tilde{\theta}^{a (2)}_s- \tilde{\theta}^{a (3)}\,,
\label{Initial-Exp-3order-St}
\ee
where all  terms on the right-hand side are expressed in terms of $\theta^a_o$. Inserting the above results 
(\ref{Theta1-lead})-(\ref{w(3)}), and performing a series of algebraic manipulations, we eventually arrive at the following simple expression:
 \bea 
(\theta^{a (3)}_{s})_{\rm Leading}&=&-\frac{2}{\eta_o-\eta_s} \int_{\eta_s}^{\eta_o}d\eta' \frac{\eta'-\eta_s}{\eta_o-\eta'} \hat{\gamma}_0^{a b}
\left[(\theta^{c (2)})_{\rm Leading}(\eta') \partial_c \partial_b \psi(\eta')
\right.
\nonumber \\
&& \left.
+\frac{1}{2} \theta^{c (1)}(\eta') \theta^{d (1)}(\eta')
\partial_c \partial_d \partial_b \psi(\eta')\right],
\label{LeadStorder3}
\eea
where the first and second-order leading expressions for $ \theta^{a (1)}$, $ \theta^{a (2)}$ are given by Eqns. (\ref{thetatilde1order_Inverted}) and (\ref{LeadStorder2}),  respectively.

Let us conclude this section by including the contribution of the Bardeen potentials to second and third order in the calculation of the leading
lensing contribution to $\theta^{a}_{s}$.
This can be simply done substituting $\psi$ with $\psi+(1/4) \left(\psi^{(2)}+\phi^{(2)}\right)+(1/12) \left(\psi^{(3)}+\phi^{(3)}\right)$
inside the leading result of Eqs. (\ref{thetatilde1order_Inverted}), (\ref{LeadStorder2}) and (\ref{LeadStorder3}), and expanding up to the third perturbative order.
We then easily obtain the following full leading results at second and third order
\bea
(\theta^{a (2)}_{s})_{\rm Full\,Leading}&=&
-\frac{1}{2}\frac{1}{\eta_o-\eta_s} \int_{\eta_s}^{\eta_o} d\eta' \frac{\eta'-\eta_s}{\eta_o-\eta'}\hat{\gamma}_0^{a b}\partial_b\left[\psi^{(2)}(\eta')+\phi^{(2)}(\eta')
\right] \nonumber \\
& &
-\frac{2}{\eta_o-\eta_s} \int_{\eta_s}^{\eta_o}d\eta' \frac{\eta'-\eta_s}{\eta_o-\eta'} \hat{\gamma}_0^{a b} 
\hat{\gamma}_0^{c d} \partial_d \psiP(\eta')
\partial_c \partial_b \psi(\eta')
\label{LeadStorder2withHO}
\eea
and
\bea 
(\theta^{a (3)}_{s})_{\rm Full\,Leading}&=&
-\frac{1}{6}\frac{1}{\eta_o-\eta_s} \int_{\eta_s}^{\eta_o} d\eta' \frac{\eta'-\eta_s}{\eta_o-\eta'}\hat{\gamma}_0^{a b}\partial_b\left[\psi^{(3)}(\eta')+\phi^{(3)}(\eta')
\right] \nonumber \\
& &
-\frac{1}{2}\frac{1}{\eta_o-\eta_s} \int_{\eta_s}^{\eta_o}d\eta' \frac{\eta'-\eta_s}{\eta_o-\eta'} \hat{\gamma}_0^{a b} 
\hat{\gamma}_0^{c d} \partial_d \psiP(\eta')
\partial_c \partial_b \left[\psi^{(2)}(\eta')+\phi^{(2)}(\eta')\right]
\nonumber 
\\
& &
-\frac{2}{\eta_o-\eta_s} \int_{\eta_s}^{\eta_o}d\eta' \frac{\eta'-\eta_s}{\eta_o-\eta'} \hat{\gamma}_0^{a b}
\left[(\theta^{c (2)})_{\rm Full\,Leading}(\eta') \partial_c \partial_b \psi(\eta')
\right.
\nonumber \\
&& \left.
+\frac{1}{2} \theta^{c (1)}(\eta') \theta^{d (1)}(\eta')
\partial_c \partial_d \partial_b \psi(\eta')\right]\,.
\label{LeadStorder3withHO}
\eea
Furthermore, anisotropic stress can be taken into account in this limit by simply replacing $\psi$ with $(\psi+\phi)/2$ as discussed in 
 \cite{Marozzi:2014kua}.
 
 We conclude this Section by recalling that we have neglected tensor and vector perturbations, in despite of the fact they are inevitably generated from scalar perturbations at second (and higher) order.
The lensing associated with tensor and vector perturbations  induced at second order  was already discussed in \cite{3,DiDio:2012bu}. In our context we
want  to underline that their contribution to  lensing   is  subleading with respect to the scalar one since it always involves, order by order, a smaller number  of transverse derivatives. In other words, our final results for the leading lensing terms are unaffected by vector and tensor perturbations while the full results are not.

\section{Expressing PG quantities directly in terms of GLC coordinates: a more direct approach}
\label{Sec4}
\setcounter{equation}{0}

We will now present a different approach, based on the direct computation of the PG angles $\theta_s^a$  (and possibly other PG quantities) as functions of the observation angles $\theta_o^a$ (which, as  already stressed, coincide with the angular coordinates $\tt^a_s$ of the GLC frame) and of the $\tau$ coordinate of the source. 
An important advantage of this new method is that, in order to give the  result in terms of just physical observables, it is sufficient to express $\tau_s$ in the final expressions in terms of redshift and GLC angles. Given the novelty of this method, our discussion here will be more detailed than the one in Sect. \ref{Sec3}.

\subsection{The coordinate transformation}
\label{4.1}

We start with the transformation expressing the PG metric  (\ref{8}) in terms of the 
GLG metric (\ref{GLCmetric}), and given in general by
\beq 
g_{\rm PG}^{\mu\nu}(y) ={\pa y^\mu \over \pa x^\a}{\pa y^\nu \over \pa x^\b}\,g_{\rm GLG}^{\a\b}(x)~,
\label{5}
\eeq
where we have denoted by $y^{\mu} = \left(\eta, \ep, \theta^a \right)$ the PG coordinates and by $x^{\a} = ( \tau, w, \ti\theta^a)$ the GLG ones.
We also recall that, if we neglect higher order corrections, the standard FLRW geometry is represented in the GLG parametrization by the following zeroth-order metric components \cite{2,3}:
\beq
(\U^{-1})^{(0)}= a^{-1}, ~~~~~~~~~ {U^{a}}^{(0)}=0, ~~~~~~~~~ {\ga^{ab}}^{(0)} = a^{-2} \ga_0^{ab},
\label{17}
\eeq
and that, to this order, the GLG and PG coordinates are connected according to Eq. (\ref{5}) by the (almost trivial) transformation:
\beq
 \eta^{(0)}= \int^\tau_{\tau_{in}} {d \tau'\over a(\tau')}, ~~~~~~~~~~\Ep{0}= w, ~~~~~~~~~~ {\theta^a}^{(0)}= \tta.
\label{18}
\eeq

Let us now include perturbations up to the desired order, and let us expand 
the PG coordinates $y^\mu$ in terms of the GLC ones around the above lowest order expression as:
\beq
y^\mu= (y^\mu)^{(0)} (x) +(y^\mu)^{(1)} (x) +(y^\mu)^{(2)} (x)  + \cdots ~.
\label{10}
\eeq 
Consequently, we expand around $(y^\mu)^{(0)}$ all components of the PG metric (\ref{8}) as
\beq
g_{\rm PG}^{\mu\nu}(y) = g_{\rm PG}^{\mu\nu}(y^{(0)}(x)) + \left({\partial g_{\rm PG}^{\mu\nu}}\over \partial y^{\rho} \right)_{y=y^{(0)}}(y^\rho)^{(1)} (x) + \cdots ~.
\label{ExpansionPGmetric}
\eeq
 Finally, we perform similar expansions, up to the required order, for all components of the (inverse) GLG metric (\ref{GLCmetric}):
\beq
\U^{-1}(x)= (\U^{-1})^{(0)}+(\U^{-1})^{(1)}+\cdots , 
\nonumber
\eeq
\beq
U^{a}(x)= {U^{a}}^{(0)}+{U^{a}}^{(1)}+\cdots , ~~~~~~
\ga^{ab}(x)= {\ga^{ab}}^{(0)}   + {\ga^{ab}}^{(1)} + \cdots
\label{16}
\eeq

Here we are interested, in particular, in the transformation providing the  angular coordinates of the PG frame computed up to second perturbative order.
To this purpose we need the first-order transformation of all PG coordinates $\eta$, $\ep$ and $\theta^a$. Hence, let us start  applying and expanding Eq. (\ref{5}) by including, for the moment, only first-order corrections. 

In this approximation, the computation of the $g_{\rm PG}^{00}$ component of Eq. (\ref{5}) then gives the following differential equation for $\eta^{(1)}$:
\beq
\pa_\tau \eta^{(1)} =- {\psi\over a} -{1\over a} \pa_w \eta^{(1)} - H \eta^{(1)}\,,
\label{19}
\eeq
where we have defined $H=\partial_\tau a/a$.
In a similar way, the computation of the transformation relative to the 
 $g_{\rm PG}^{++}$ component of Eq. (\ref{5}) gives the  differential equation for $\eta^{+(1)}$:
\beq
\pa_\tau \Ep{1} =- {2\psi\over a} ,
\label{20}
\eeq
and the transformation of the 
 $g_{\rm PG}^{+a}$ component of Eq. (\ref{5}) gives the  differential equation for $\theta^{a(1)}$:
\beq
\pa_\tau {\theta^a}^{(1)} ={1\over a} \ga_0^{ab} \pa_b \Ep{1} \,,
\label{21}
\eeq
where we recall that $\pat_b$ denotes partial derivative with respect to $\tilde{\theta}^a$. 
It should be noted that, according to the expansion (\ref{ExpansionPGmetric}), all  components of the PG metric (such as $a, \psi$) appearing here and in the subsequent equations are always functions of the unperturbed PG coordinates 
$(y^\mu)^{(0)}$, namely, considering the relation given in Eq. (\ref{18}), they are functions of the GLC coordinates $(\tau, w, \tilde{\theta}^a)$.

Considering the same boundary conditions as those used in Sec. \ref{Sec3}, 
integrating with respect to $\tau$ the last two equations, and using the fact that $w$ and $\tt^a$ are constant along a null geodesic, we easily obtain, along such a curve,

\beq
\eta^{+(1)}
= 2 \int^{\tau_o}_{\tau} {d \tau'\over a(\tau')} \,\psi (\tau', w, \tt ),
\label{22}
\eeq
and the first-order expression for the deflection angle in terms of GLC coordinates is then given by:
\beq
\theta^{a(1)}
= -2 \int_\tau^{\tau_o} {d \tau'\over a(\tau')} \,\ga_0^{ab} (\tau' , w, \tt)
\int_{\tau'}^{\tau_o} {d \tau''\over a(\tau'')} \,\pat_b
\psi (\tau'', w, \tt ).
\label{23}
\eeq
Let us also report, for later use, the first-order result for the conformal-time coordinate $\eta^{(1)}$. The general solution of Eq. (\ref{19}) gives
\beq
\eta^{(1)} 
= -{1\over a} \int^\tau_{\tau_{in}} d \tau' \,  \psi \left(\tau', \xi(w,\tau, \tau'), \tt \right),
\label{24}
\eeq
where
\beq 
\xi(w,\tau, \tau')= w- \eta^{(0)}(\tau)+\eta^{(0)}(\tau'),
\label{25}
\eeq
as can be easily checked by differentiating Eq. (\ref{24}) and inserting the results into Eq. (\ref{19}). 

In order to compute the full second-order angular transformation, $\theta^{a(2)}$, we need also the expressions for the first-order perturbations of the GLG metric (\ref{GLCmetric}). They can be obtained, as before, from the components of the general transformation (\ref{5}). Considering in particular the transformation of the 
$g_{\rm PG}^{+0}$ component we obtain, to first order, 
\beq
(\U^{-1})^{(1)}= -{1\over a} \pa_w \Ep{1}- \pa_\tau \eta^{(1)} -2\,H\eta^{(1)},
\label{26}
\eeq
or, using Eq. (\ref{19}):
\beq
(\U^{-1})^{(1)}= -{1\over a} \pa_w \Ep{1} + {\psi\over a} +{1\over a} \pa_w \eta^{(1)}-H\eta^{(1)},
\label{27}
\eeq
Similarly, considering the transformation of the $g_{\rm PG}^{0a}$ component, we obtain
\beq
{U^{a}}^{(1)}= \ga_0^{ab} \pat_b \eta^{(1)} - a  \pa_\tau \th{1}{a} - \pa_w \th{1}{a} ,
\label{28}
\eeq
or, using Eq. (\ref{21}),
\beq
{U^{a}}^{(1)}= \ga_0^{ab} \pat_b \eta^{(1)} - \ga_0^{ab}  \pat_b \Ep{1} - \pa_w \th{1}{a}  .
\label{29}
\eeq
Finally, from the $g_{\rm PG}^{ab}$ component of Eq. (\ref{5}):
\bea
{\ga^{ab}}^{(1)} &=& {2\psi\over a^2} \ga_0^{ab} -{\ga_0^{bc}\over a^2}
\pat_c \th{1}{a}-{\ga_0^{ac}\over a^2}
\pat_c \th{1}{b} - {2\over a} \ga_0^{ab}
\,H \eta^{(1)}   \nonumber \\
&&- {2\over a^2} \ga_0^{ab}
{ \Ep{1}-\eta^{(1)} \over \Ep{0}-\eta^{(0)}}
+{\pat_c\gamma_0^{ab}\over a^2}\theta^{c(1)}.
\label{30}
\eea

We are now in the position of extending to second order the transformation of the angular coordinates. We must compute, first of all, the differential equation determining to second order the light-cone coordinate $\ep$. Considering the $g_{\rm PG}^{++}$ component of Eq. (\ref{5}), and including all second-order contributions, we obtain the differential condition:
\bea
\pa_\tau \Ep{2} &=& -{1\over 2a} \left( \psi^{(2)} + \vp^{(2)} \right) +{1\over 2a} \ga_0^{ab} \pat_a \Ep{1}\pat_b \Ep{1}
+{2\psi \over a}\pa_w \eta^{(1)}
+2\,H\psi  \eta^{(1)}
\nonumber \\
&&
 -2 \,\eta^{(1)}\, \pa_\tau \psi - {2\over a} \Ep{1} \pa_w \psi
-{2\over a}\, \th{1}{a} \pat_a \psi \,,
\label{31}
\eea
where we have used Eq. (\ref{27}) for $(\U^{-1})^{(1)}$, and Eq. (\ref{20}) for $\pa_\tau \Ep{1}$. By integrating with respect to $\tau$ along a null geodesic, as before, and using the first-order results of Eqs. (\ref{22})--(\ref{24}) for $\eta$, $\ep$ and $\theta^a$, we thus obtain (hereafter, for simplicity, we omit to write the dependence of $\ga_0$ and $\psi$  on the constant  arguments $w$ and $\tt^a$):
\bea
\eta^{+(2)}
&=&{1\over 2} \int^{\tau_o}_\tau {d \tau'\over a(\tau')} \left( \psi^{(2)}(\tau')+ \vp^{(2)}(\tau')\right)
\nonumber \\ &&
-2 \int^{\tau_o}_\tau {d \tau'\over a(\tau')}\ga_0^{ab}(\tau') \left(\int_{\tau'}^{\tau_o} {d \tau''\over a(\tau'')}  \pat_a \psi(\tau'')\right) \left(\int_{\tau'}^{\tau_o} {d \tau'''\over a(\tau''')}  \pat_b \psi(\tau''')\right)
\nonumber \\ &&
+2\int_\tau^{\tau_o} {d \tau'\over a^2(\tau')}\, \psi(\tau') \int^{\tau'}_{\tau_{in}} d \tau'' \pa_w \psi\left( \tau'', \xi(w, \tau', \tau'') \right)
\nonumber \\ &&
+2\int_\tau^{\tau_o} {d \tau'\over a(\tau')}\, H(\tau')\,\psi(\tau') \int^{\tau'}_{\tau_{in}} d \tau'' \, \psi\left( \tau'', \xi(w, \tau', \tau'') \right)
\nonumber \\ &&
-2\int_\tau^{\tau_o} {d \tau'\over a(\tau')}  \left( \pa_{\tau'} \psi(\tau')\right)\int^{\tau'}_{\tau_{in}} d \tau'' \,\psi\left( \tau'', \xi(w, \tau', \tau'') \right)
\nonumber \\ &&
+4\int_\tau^{\tau_o} {d \tau'\over a(\tau')}  \left( \pa_w \psi(\tau')\right) \int_{\tau'}^{\tau_o}  {d \tau''\over a(\tau'')}  \,\psi \left( \tau'' \right)
\nonumber \\ &&
-4 \int_\tau^{\tau_o} {d \tau'\over a(\tau')}\left(\pat_a \psi(\tau')\right) \int_{\tau'}^{\tau_o} {d \tau''\over a(\tau'')}  \ga_0^{ab}(\tau'')
\int_{\tau''}^{\tau_o} {d \tau'''\over a(\tau''')}  \pat_b \psi(\tau''')\,.
\label{32}
\eea

Finally, considering the transformation of the $g_{\rm PG}^{+a}$ component of Eq. (\ref{5}), and using the results of  Eqs. (\ref{27}), (\ref{28}) and  (\ref{30}) for the first-order components of the GLG metric, we obtain, to second-order:
\bea
\pa_\tau \th{2}{a} &=& 
{1\over a} \ga_0^{ab}\pat_b \Ep{2} 
-{1\over a} \ga_0^{ac} \pat_b \Ep{1} \pat_c \th{1}{b} 
+{1\over a}  \pat_b \Ep{1} \th{1}{c} \partial_c  \ga_0^{ab} \nonumber \\
& & 
+{1\over a} \ga_0^{ab} \Bigg[
+2 \psi \pat_b \eta^{(1)} + \psi  \pat_b \Ep{1}
-  \pat_b \Ep{1}  \pa_w \eta^{(1)} 
\nonumber \\
& & 
- aH\, \eta^{(1)}\,\pat_b \Ep{1} -2  \,
{ \Ep{1}-\eta^{(1)} \over  \Ep{0}-\eta^{(0)}} \,\pat_b \Ep{1}
\Bigg].
\nonumber \\
\label{Deflection-2order-GLC}
\eea
By inserting the first-order results (\ref{22})--(\ref{24}) for $\eta$, $\ep$ and $\theta^a$, the second-order result (\ref{32}) for $\eta^{+(2)}$, and integrating with respect to $\tau$, one immediately obtains from the above equation 
the second-order contribution to the sought transformation, expressing the PG angular coordinates $\theta^a$ in terms of the GLG coordinates:
\beq
\theta^{a(2)} (\tau, w, \tt^b) =-\int_\tau^{\tau_o} d\tau' \left( \pa_\tau \theta^a \right)^{(2)}  (\tau', w, \tt^b).
\label{422}
\eeq

\subsection{Light deflection to second order}
\label{4.3}

In order to reproduce the results of the previous Section we need to express the PG coordinates of the source, $\theta_s^a$, in terms of both the observation angles $\theta_o^a$ (which, as already stressed, exactly coincides with the GLG angles $\tt_s^a$) and of the observed redshift $z_s$.

Let us start considering, as in Sect. \ref{Sec32}, the leading second-order part of the coordinate transformation corresponding to the lensing contributions to $\theta_s^{a(2)}$ generated by three angular derivatives, and contained inside the first three terms of Eq. (\ref{Deflection-2order-GLC})~\footnote{
As in Sec. \ref{Sec3}, the contribution arising from writing the deflection angles in terms 
of the observed redshift are subleading with respect to the leading lensing contribution.}. Let us call them $\theta^{a (2)}_{s,B} $. Using Eqs.  (\ref{22}),  (\ref{23}) for  $\eta^{+(1)}$ and $\theta^{a(1)}$, and considering the second and the last line of Eq.  (\ref{32}) for the leading terms of  $\eta^{+(2)}$, we obtain:
\bea
&&
\theta^{a (2)}_{s,B}  = \nonumber  \\
&&
2 \int^{\tau_o}_{\tau_s} {d \tau'\over a(\tau')} \ga_0^{ab} (\tau') \int^{\tau_o}_{\tau'} {d \tau''\over a(\tau'')}\pat_b \left[\ga_0^{cd}(\tau'') \left( \int^{\tau_o}_{\tau''} {d \tau'''\over a(\tau''')} \pat_c \psi (\tau''') \right) \left( \int^{\tau_o}_{\tau''} {d \tau'''\over a(\tau''')} \pat_d \psi (\tau''') \right)\right]
\nonumber \\ &&
+4 \int^{\tau_o}_{\tau_s} {d \tau'\over a(\tau')} \ga_0^{ab} (\tau') \int^{\tau_o}_{\tau'} {d \tau''\over a(\tau'')}\pat_b \left[\pat_c \psi (\tau'')  \int^{\tau_o}_{\tau''} {d \tau'''\over a(\tau''')}\ga_0^{cd}(\tau''')
\int^{\tau_o}_{\tau'''} {d \tau''''\over a(\tau'''')}\pat_d \psi(\tau'''') \right]
\nonumber \\ &&
-4 \int^{\tau_o}_{\tau_s} {d \tau'\over a(\tau')} \ga_0^{ac} (\tau') \left( \int^{\tau_o}_{\tau'} {d \tau''\over a(\tau'')}\pat_b \psi (\tau'') \right)
\left[\int^{\tau_o}_{\tau'} {d \tau''\over a(\tau'')}\pat_c \left( \ga_0^{bd} \int^{\tau_o}_{\tau''} {d \tau'''\over a(\tau''')}\pat_d \psi(\tau''') \right) \right]
\nonumber \\ &&
+4 \int^{\tau_o}_{\tau_s} {d \tau'\over a(\tau')} \left(\partial_c \ga_0^{ab} (\tau')\right) \left( \int^{\tau_o}_{\tau'} {d \tau''\over a(\tau'')}\pat_b \psi (\tau'') \right)
\left[\int^{\tau_o}_{\tau'} {d \tau''\over a(\tau'')}\ga_0^{cd} \int^{\tau_o}_{\tau''} {d \tau'''\over a(\tau''')}\pat_d \psi(\tau''') \right]. 
\nonumber  \\
\eea

Let us now go from the GLC coordinates $(\tau,w)$ to the zeroth-order PG coordinates $(\eta^{(0)},\eta^{+(0)})$ using Eq.(\ref{18}).
Being already at second order, we can then drop the suffix $(0)$ and move to the standard PG coordinates $(\eta,r)$  considering the zeroth-order relation $w=\eta+r= w_o= \eta_o$.

After several algebraic manipulations and integrations by parts we can then write the above result in the equivalent, more useful  form:
\bea
\theta^{a (2)}_{s,B}&=&-\frac{2}{\eta_o-\eta_s}\int_{\eta_s}^{\eta_o}d\eta'\frac{\eta'-\eta_s}{\eta_o-\eta'}\hat\ga_0^{ac}\hat\ga_0^{bd}\pa_d\psiP(\eta')
\pa_c\pa_b\psi(\eta')
\nonumber\\
&+&\frac{2}{\eta_o-\eta_s}\int_{\eta_s}^{\eta_o}d\eta'\frac{\eta'-\eta_s}{\eta_o-\eta'}\left[ \hat\ga_0^{bc}\pa_b\psi(\eta')\left( \pa_c\hat\ga_0^{ad} \right)\pa_d\psiP(\eta')
-\hat\ga_0^{ad}\pa_b\psi(\eta')\left( \pa_d\hat\ga_0^{cb} \right)\pa_c\psiP(\eta') \right]\nonumber\\
&+&\frac{4}{\eta_o-\eta_s}\int_{\eta_s}^{\eta_o}d\eta'\frac{\eta'-\eta_s}{\eta_o-\eta'}\left[ -\ga_0^{bc}\left(\pa_b\g{\eta'}{\eta''}\right)\left( \pa_c\ga_0^{ad} \right)\left(\pa_d \g{\eta'}{\eta''} \right)\right.\nonumber\\
&+&\left.\frac{1}{2}\ga_0^{ad}\left( \pa_d\hat\ga_0^{bc} \right)\left(\pa_c\g{\eta'}{\eta''}\right)\left(\pa_b \g{\eta'}{\eta''} \right)\right]\nonumber\\
&+&4\int_{\eta_s}^{\eta_o}d\eta' \left[\left(  \pa_b\g{\eta'}{\eta''} \right)\left( \pa_d\ga_0^{ab} \right)\int_{\eta'}^{\eta_o}d\eta''\ga_0^{dc}\left( \pa_c\g{\eta''}{\eta'''} \right) \right],
\label{thetaB}
\eea
where we have used Eq. (\ref{Lensing-Potential}) for the lensing potential.
Neglecting all  subleading terms without three angular derivatives acting on the metric fluctuations we are left only with the first-line contribution, and thus we recover for $(\theta^{a (2)}_s)_{\rm Leading}$ exactly the same result as the one given by the method of the previous section, see Eq. (\ref{LeadStorder2}) (also in agreement with the results presented in \cite{Bonvin:2015uha}). 

We also note that, by applying the following identity, 
\bea
&&
\left( \pa_b\hat\ga_0^{ad} \right)\hat\ga_0^{bc}\pa_d\psiP\pa_c\psiP=
\nonumber \\ 
&=&
4\int_{\eta_s}^{\eta_o}d\eta'\left( \pa_d\g{\eta'}{\eta''} \right)\left( \pa_b\ga_0^{ad} \right)\int_{\eta'}^{\eta_o}d\eta''\ga_0^{bc}\left( \pa_c\g{\eta''}{\eta'''} \right)\nonumber\\
&+&4\int_{\eta_s}^{\eta_o}d\eta'\ga_0^{bc}\left( \pa_c\g{\eta'}{\eta''}\right)\int_{\eta'}^{\eta_o}d\eta''\left(\pa_b \ga_0^{ad} \right)\left(\pa_d \g{\eta''}{\eta'''} \right),
\eea
it can be explicitly checked  that the result of Eq. (\ref{thetaB})  for $\theta^{a (2)}_{s,B}$  exactly coincides with the result for $\theta^{a (2)}_{s,A}$ given 
in Eq. (\ref{theatas-S-A-form1-simplified}), i.e. that $\theta^{a (2)}_{s,A}\equiv \theta^{a (2)}_{s,B}$.

Let us now provide the full second-order result for $\theta_s^a$ not only in terms of the observation angle $\theta_o^a$ but also in terms of the observed redshift $z_s$, related to the coordinate $\tau_s$ by the exact expression 
(\ref{redshift}). To this purpose, similarly to what done in Sec. \ref{Sec3} and in previous papers  \cite{2,3,Marozzi:2014kua}, we expand $\tau_s$ around the convenient values $\bar \tau_s^{(0)}$ defined in such a way that the observed redshift $z_s$ is given by\footnote{Another computational advantage of the approach of this section is that we  need to replace only one model-dependent parameter, $\tau_s$, with the observable quantity $z_s$, instead of the two parameters $\eta_s$ and $r_s$ appearing within the approach of Sect. \ref{Sec3}.}:
\beq
1+z_s={\Ups(\tau_o, w_o, \tt_o)\over \Ups (\tau_s, w_o, \tt_s)} \equiv \frac{a(\tau_o)}{a(\bar{\tau}^{(0)}_s)}\,.
\label{Fiducial-model-GLC-coordinates}
\eeq
By setting $\tau_s= \bar{\tau}^{(0)}_s+\bar{\tau}^{(1)}_s+ \cdots$, by using for $\Ups^{-1}$ the first-order result  (\ref{27}), and expanding to first order the above equation around $\bar{\tau}^{(0)}_s$ we easily obtain
\beq
{\Ups(\tau_o)\over \Ups (\tau_s)} = \frac{a(\tau_o)}{a(\bar{\tau}^{(0)}_s)} \left[ 1 +  \psi - \pa_w \eta^{+(1)}+ \pa_w \eta^{(1)} 
- a H \eta^{(1)} -H \bar{\tau}^{(1)}_s 
+ \cdots \right]_{\bar{\tau}^{(0)}_s},
\eeq
from which, by applying Eq. (\ref{Fiducial-model-GLC-coordinates}), we have
\beq
\bar{\tau}^{(1)}_s= \frac{1}{H} \left[\psi -
 \pa_w \Ep{1} + \pa_w \eta^{(1)}-
aH\, \eta^{(1)} \right]\,,
\label{tau1-OR}
\eeq
where  everything is now evaluated at $\bar{\tau}^{(0)}_s$.
Notice that, following Sect. \ref{Sec32}, we have neglected already at this stage the perturbative contributions evaluated at the observer position.

The final expression for $\theta_s^a$ in terms of the observation variables $(\theta_o^a, z_s)$ can now be obtained by Taylor expanding the angular coordinate transformation around $\bar{\tau}^{(0)}_s$ .  Including all contributions up to second order, using the fact that $w$ and $\tilde{\theta}^a$ are constant along the null ray trajectory and that, to zeroth order,  
 $\theta^{a(0)}_s =\tt^a_s  =\theta^a_o$, 
we can write:
\bea
\theta^a_s (z_s, \theta_o)&=& \theta^a_o + \theta^{a(1)}  (\bar{\tau}_s^{(0)} , w, \theta_o) +  \theta^{a(2)} (\bar{\tau}_s^{(0)}, w, \theta_o) 
\nonumber \\ &+&
\bar{\tau}_s^{(1)}  (\pa_\tau \theta^a)^{(1)}  (\bar{\tau}_s^{(0)}, w, \theta_o) 
+ \cdots \,.
\label{231}
\eea
Here $\theta^{a(1)}$, $\theta^{a(2)}$ are given, respectively, by the expressions (\ref{23}), (\ref{422}) evaluated (like $\bar{\tau}_s^{(1)} $) at $\tau_s= \bar \tau_s^{(0)}$. 
The time parameter $\bar\tau_s^{(0)}$, in its turn, is directly related to the observed redshift $z_s$  and to the parameter $\bar{\eta}_s^{(0)}$ of Sect. \ref{Sec32} by  the standard relation $d \bar{\tau}_s^{(0)}/a= d\bar{\eta}_s^{(0)}= - dz_s(1+z_s)^{-1} {\mathcal{H}_s}^{-1}$ (where ${\mathcal{H}_s}=(\pa_\eta a /a)_{\bar{\eta}_s^{(0)}}$, as before). 

By taking into account all contributions we can now perform some useful simplification in the second-order part of Eq. (\ref{231}). Switching to the conformal time parameter $\bar{\eta}_s^{(0)}$ defined above, 
and separating the first and second order contributions, we then find: 
\bea
 \theta^{a (1)}_s (\theta^a_o, z_s)
&=&  -2 \int_{\bar{\eta}^{(0)}_s}^{\eta_o}d\eta' \ga_0^{ab} \int_{\eta'}^{\eta_o}d\eta'' \pa_b  \psi (\eta'');  
\\
 \theta^{a (2)}_s (\theta^a_o, z_s)
&=& 
\theta^{a (2)}_{s,B}
-2\int_{\bar\eta_s^{(0)}}^{\eta_o} d\eta'  \psi(\eta')\ga_0^{ab} \pa_b\g{\eta'}{\eta''}
-2\int_{\bar\eta_s^{(0)}}^{\eta_o}d\eta'\ga_0^{ab}\pa_b\int_{\eta'}^{\eta_o}d\eta''\psi^2(\eta'')
\nonumber
\\
& &\!\!\!\!\!\!\!\!\!\!\!\!\!\!\!\!\!\!\!\!\!\!\!\!\!-\frac{2}{a\Hcal_s}\ga_0^{ab}\left(\pa_b\g{\eta_s}{\eta'}\right)\pa_w\int_{\eta_{in}}^{\bar\eta_s^{(0)}}d\eta'a(\eta')\psi\left(\eta',\xi(\eta_s,\eta')\right)\nonumber\\
& &\!\!\!\!\!\!\!\!\!\!\!\!\!\!\!\!\!\!\!\!\!\!\!\!\!+\int_{\bar\eta_s^{(0)}}^{\eta_o}d\eta'\ga_0^{ab}\pa_b\left\{-\frac{1}{2}\int_{\eta'}^{\eta_o}d\eta''\left( \psi^{(2)}+\phi^{(2)} \right)
-4\int_{\eta'}^{\eta_o}d\eta''\left[ \left(\pa_w\psi(\eta'')\right)\g{\eta''}{\eta'''} \right] \right\}\nonumber\\
& &\!\!\!\!\!\!\!\!\!\!\!\!\!\!\!\!\!\!\!\!\!\!\!\!\!-\int_{\bar\eta_s^{(0)}}^{\eta_o}d\eta'\ga_0^{ab}\left[ 2\psi(\eta')\pa_b\g{\eta'}{\eta''}
-\frac{8}{\eta_o-\eta'}\g{\eta'}{\eta''}\pa_b\g{\eta'}{\eta''}\right]\nonumber\\
& &\!\!\!\!\!\!\!\!\!\!\!\!\!\!\!\!\!\!\!\!\!\!\!\!\!+4\int_{\bar\eta_s^{(0)}}^{\eta_o}d\eta'\left\{\left( \pa_b\g{\eta'}{\eta''}\right)\left(\pa_d \ga_0^{ab} \right)\int_{\eta'}^{\eta_o}d\eta''\ga_0^{dc}\pa_c\g{\eta''}{\eta'''} \right\}\nonumber\\
& &\!\!\!\!\!\!\!\!\!\!\!\!\!\!\!\!\!\!\!\!\!\!\!\!\!+\frac{2}{\Hcal_s}\left[ \psi_s-2\,\pa_w\g{\eta_s}{\eta'} \right]\ga_0^{ab}\pa_b\g{\eta_s}{\eta'}\,.
 \label{233}
 \eea
 
 We can immediately notice that the first order result exactly coincide with the one previously obtained in Sect. \ref{Sec3}, see Eq. (\ref{FinalThetaStorder1}). In addition, using the fact that $\theta^{a (2)}_{s,B}=\theta^{a (2)}_{s,A}$, performing further simplifications by noticing, for instance, that
\bea
& & \!\!\!\!\!\frac{1}{a} \pa_w\int_{\eta_{in}}^{\bar\eta_s^{(0)}}d\eta'a(\eta')\psi\left(\eta',\xi(\eta_s,\eta')\right)=
\int_{\eta_{in}}^{\bar{\eta}_s^{(0)}} d\eta' \frac{a(\eta')}{a(\eta_s)}\partial_r \psi\left(\eta', \eta_s-\eta_o, \theta^a_o\right)=v_{||s}\,,
\\
& &\!\!\!\!\int^\eta d \eta' \pa_w \psi(\eta', w, \theta) = \int^\eta d \eta' \left(-{d \psi\over d \eta'} + \pa_{\eta'} \psi \right)=
- \psi(\eta) +\int^\eta d \eta' \pa_{\eta'} \psi, 
\label{234}
\eea
and integrating by part several times, it can be eventually shown that the result of  Eq. (\ref{233}) is exactly equivalent to the full expression for $\theta_s^{a(2)}$ obtained with the complementary approach of Sect. \ref{Sec3}, and already  reported in Eq. (\ref{Final-theta2-general}).

\subsection{Leading third-order contributions}
\label{4.4}

Following the same procedure of Sect.  \ref{4.1} we  now extend to third order the computation of the angular coordinate
$\theta_s^a$ as a function of $\theta_o^b$, considering however only the contributions of the leading lensing terms with five angular derivatives of the metric fluctuations. As already stressed, in this leading approximation there is no need of performing also the  redshift expansion. Considering the  component $g^{+a}_{\rm PG}$ of the coordinate transformation (\ref{5}) we then obtain  the following differential equation for $\theta^{a(3)}$:
\bea
\left(\pa_\tau\th{3}{a} \right)_{\rm Leading}&=& {1\over a}\ga_0^{ab}\left(\pat_b \Ep{3}\right)_{\rm Leading}
+a \left({\ga^{ab}}^{(2)}\right)_{\rm Leading} \pat_b \Ep{1}
\nonumber \\
&& + {\ga_0^{bc}\over a}  \left(\pat_c \th{2}{a}\right)_{\rm Leading} \pat_b \Ep{1} -  {\ga_0^{ac}\over a}
 \pat_c \th{1}{b} \left(\pat_b \Ep{2}\right)_{\rm Leading} 
\nonumber \\
&&-{1\over a} \left[\ga_0^{bd} \pat_d \th{1}{c}
+\ga_0^{cd} \pat_d \th{1}{b}
\right]
\pat_c \th{1}{a} \pat_b \Ep{1}.
\label{41}
\eea
We can use the results of the previous sections for $\theta^a$ and $\eta^+$ to the first and second order, but we still have to compute
$\Ep{3}$ and ${\ga^{ab}}^{(2)}$.

Let us start with ${\ga^{ab}}^{(2)}$. By expressing the component $g_{\rm PG}^{ab}$ of the transformation (\ref{5}) up to second order, by using Eqs. (\ref{8}) and (\ref{9}), and keeping only leading lensing terms with four angular derivatives, we obtain:
\bea
\label{42}
 \left({\ga^{ab}}^{(2)}\right)_{\rm Leading}&=&
-\frac{\ga_0^{bc}}{a^2}\left(\pat_c\th{2}{a}\right)_{\rm Leading}
-\frac{\ga_0^{ac}}{a^2}\left(\pat_c\th{2}{b}\right)_{\rm Leading}
+\frac{\gamma_0^{bd}}{a^2}\pat_d\th{1}{c}\pat_c\th{1}{a}
\nonumber\\
&& +\frac{\gamma_0^{ad}}{a^2}\pa_d\th{1}{c}\pat_c\th{1}{b}
+\frac{\gamma_0^{cd}}{a^2}\pat_c\th{1}{a}\pat_d\th{1}{b}.
\eea
The above equation (\ref{41}) for $(\theta^a)^{(3)}$ can thus be rewritten as
\bea
\label{43}
\left(\pa_\tau\th{3}{a} \right)_{\rm Leading}&=&
\frac{\ga_0^{ab}}{a}\left(\pat_b \Ep{3}\right)_{\rm Leading}
+\frac{\gamma_0^{ad}}{a}\pat_d\th{1}{c}\pat_c\th{1}{b}\pat_b\Ep{1}
\nonumber\\
&&-\frac{\ga_0^{ac}}{a} \left[
\pat_c\th{1}{b}\left(\pat_b\Ep{2}\right)_{\rm Leading}
+\left(\pat_c\th{2}{b}\right)_{\rm Leading}\pat_b\Ep{1}\right].
\eea

Let us finally compute $(\eta^+)^{(3)}_{\rm Leading}$. To this purpose we first need the leading, third-order contributions to the component $g_{\rm PG}^{++}$ of the inverted PG metric (\ref{8}). By using Eq. (\ref{9}), and expanding to third order the  perturbation 
$\psi$ with respect to the PG coordinates (according to Eqs. (\ref{10}), (\ref{ExpansionPGmetric})),  we obtain the leading terms
\beq
\left(g_{\rm PG}^{++}\right)^{(3)}_{\rm Leading}={4\over a^2} \left(\psi^{(3)}\right)_{\rm Leading}\equiv {4\over a^2} \left(\th{2}{c}\right)_{\rm Leading} \pat_c \psi +{2\over a^2}  \theta^{c(1)}
\theta^{d(1)}\pat_c \pat_d \psi.
\label{44}
\eeq
By computing the $g_{\rm PG}^{++}$ component of the transformation (\ref{5}), and using for $(\ga^{ab})^{(1)}$ the result (\ref{30}), we then find that the differential equation for $\eta^+$ with leading, third order contributions, is given by
\bea
\left(\pa_\tau\Ep{3} \right)_{\rm Leading}&=&
\frac{\ga_0^{ab}}{a}\pat_a\Ep{1}\left(\pat_b\Ep{2}\right)_{\rm Leading}
-\frac{\ga_0^{bc}}{2a}\pat_c\th{1}{a}\pat_a\Ep{1}\pat_b\Ep{1}
\nonumber\\ &-&
\frac{\ga_0^{ac}}{2a}\pat_c\th{1}{b}\pat_a\Ep{1}\pat_b\Ep{1}
-{2\over a} \left(\th{2}{c}\right)_{\rm Leading} \pat_c \psi -{1\over a}  \theta^{c(1)}
\theta^{d(1)}\pat_c \pat_d \psi.
 \nonumber\\ &&
\eea
By integrating with respect to $\tau$, and inserting the result into Eq. (\ref{43}), we eventually arrive at the explicit form of the leading third-order contribution to the angle $\theta_s^a$.

Switching to the conformal time coordinate we can then compare such a leading expression  with the analogous 
result obtained with the complementary approach of Sect. \ref{Sec3}, and 
presented in Eq.(\ref{LeadStorder3}).  After several algebraic manipulations and integrations by parts we can  show that the two expressions exactly coincide, thus confirming the equivalence and the validity of the two approaches illustrated in Sect. \ref{Sec3} and Sect. \ref{Sec4}.

\section{Our leading-order results and and the so-called lens equation}
\label{Sec5}
\setcounter{equation}{0}
In this Section we would like to establish a connection between our (leading lensing) results and what is sometimes referred to as the (linearized) lens equation \cite{lenseq}.
This will also facilitate the comparison of our results with those in the literature.
 Let us introduce an amplification matrix $\mathcal A^a_b$ as the derivative of the angular coordinates of the source with respect to the angular direction of the light ray received at the observer's position, namely:
\be 
{\cal A}^a_b=\frac{\partial \theta_s^a}{\partial \theta_o^b}.
\label{AmpMat}
\ee
Using this definition, the results of Sect. \ref{Sec3} and \ref{Sec4} provide
  all we need for evaluating $\mathcal A^a_b$. 
  
 We want to stress, however, that the above definition of $\mathcal A^a_b$ is not fully satisfactory from the theoretical point of view. One obvious objection is that (\ref{AmpMat}) depends on the coordinate system one is using. A better way would be to start from the Jacobi map connecting a suitably projected displacement vector at the source to the angle at the observer 
(see, for instance, \cite{SEF,Lewis:2006fu,Fanizza:2014baa,Reimberg:2012uc})
and then introduce the angular coordinate  of  the source by dividing the displacement vector by an unperturbed distance between source and observer. This definition has the advantage of being gauge independent but instead depends on the  reference background model used to define $\theta_s^a$. We refer to the nice review paper \cite{Sasaki} for a thorough discussion of all the ambiguities one encounters in defining such a matrix as well as the correct equation it is supposed to obey. 

Here, more modestly, we shall adopt the definition (\ref{AmpMat}) and show that its leading lensing contributions satisfy, up to the third perturbative order,  the  above-mentioned (linearized) lens equation. 
Equivalently, by defining the deformation part of the amplification matrix by subtracting the zeroth order contribution, i.e. by introducing the convenient quantity $\Psi^a_b=\delta^a_b -  {\cal A}^a_b$, we will show that such a quantity satisfies the following equation \cite{lenseq}:
\bea 
 \Psi^a_b&=&\frac{2}{\eta_o-\eta_s}\int_{\eta_s}^{\eta_o} d \eta' \frac{\eta'-\eta_s}{\eta_o-\eta'}\, \hat{\gamma}_0^{ac}\partial_c
 \partial_d \psi(\eta',\eta_o-\eta', \theta^a) {\cal A}^d_b \nonumber \\
 &=&\frac{2}{\eta_o-\eta_s}\int_{\eta_s}^{\eta_o} d \eta' \frac{\eta'-\eta_s}{\eta_o-\eta'}\, \hat{\gamma}_0^{ac}\partial_c
 \partial_d \psi(\eta',\eta_o-\eta', \theta^a) \left[\delta^d_b-{\Psi}^d_b\right]  \,.
 \label{LensEq}
 \eea
whose structure is clearly well suited for an iterative (or in some cases perhaps even an exact) solution.

In order to check that this  lens equation reproduces our results up to third order we expand  it up to the desired $n$-th order by setting $(\Psi^a_b)^{(0)}=0$: 
\be
(\Psi^a_b)^{(n)}=-\frac{\partial \theta_s^{a (n)}}{\partial \theta_o^b}, 
~~~~~~~~~~~~ n \geq 1.
\ee
We will obtain, in this way, iterative solutions for the deformation matrix $(\Psi^a_b)^{(n)}$ and for the corresponding angular deflection 
$ \theta_s^{a (n)}$.
Stopping the computation at first order we have \cite{lenseq}, in particular,
\be 
(\Psi^a_b)^{(1)}=\frac{2}{\eta_o-\eta_s}\int_{\eta_s}^{\eta_o} d \eta' \frac{\eta'-\eta_s}{\eta_o-\eta'} \,\hat{\gamma}_0^{ac}\partial_c
 \partial_b \psi(\eta',\eta_o-\eta', \theta^a_o) \,,
\ee
which corresponds to
\be 
\theta^{a (1)}_s=-\frac{2}{\eta_o-\eta_s}\int_{\eta_s}^{\eta_o} d \eta' \frac{\eta'-\eta_s}{\eta_o-\eta'} \,\hat{\gamma}_0^{ac}\partial_c
\psi(\eta',\eta_o-\eta', \theta^a_o) \,,
\ee
and  exactly coincides with the first-order result (see e.g. Eq. (\ref{FinalThetaStorder1})). 
A second-order computation gives 
\be 
(\Psi^a_b)^{(2)}=\frac{2}{\eta_o-\eta_s}\int_{\eta_s}^{\eta_o} d \eta' \frac{\eta'-\eta_s}{\eta_o-\eta'}  \,\hat{\gamma}_0^{ac}
\left[\partial_c
 \partial_b \partial_d \psi(\eta') \theta^{d (1)}-  \partial_c \partial_d \psi(\eta') \Psi^{d (1)}_b 
 \right] \,,
\ee
which corresponds to 
\be 
\theta^{a (2)}_s=-\frac{2}{\eta_o-\eta_s}\int_{\eta_s}^{\eta_o} d \eta' \frac{\eta'-\eta_s}{\eta_o-\eta'}  \,\hat{\gamma}_0^{ac}\partial_c
\partial_d \psi(\eta')\,\theta^{d (1)}\,,
\ee
and exactly reproduces the second-order leading result of 
 Eq. (\ref{LeadStorder2}) (also in agreement with   \cite{Bonvin:2015uha}, as already stressed). 
Finally, at third order we have
\bea
(\Psi^a_b)^{(3)}&=&\frac{2}{\eta_o-\eta_s}\int_{\eta_s}^{\eta_o} d \eta' \frac{\eta'-\eta_s}{\eta_o-\eta'}  \,\hat{\gamma}_0^{ac}
\left[\partial_c
 \partial_b \partial_d \psi(\eta') \theta^{d (2)}+\frac{1}{2} \partial_c \partial_b \partial_d \partial_e \psi(\eta') \theta^{d (1)}
  \theta^{e (1)} \right.\nonumber \\ 
 & & \left.
 -\partial_c \partial_d \partial_e \psi(\eta') \theta^{e (1)}  \Psi^{d (1)}_b
   -  \partial_c \partial_d \psi(\eta') \Psi^{d (2)}_b
 \right] \,,
\eea
which corresponds to 
\be 
\theta^{a (3)}_s=-\frac{2}{\eta_o-\eta_s}\int_{\eta_s}^{\eta_o} d \eta' \frac{\eta'-\eta_s}{\eta_o-\eta'} \,\hat{\gamma}_0^{ac}
\left[\partial_c \partial_d \psi(\eta')\,\theta^{d (2)} +\frac{1}{2} \partial_c \partial_d \partial_e \psi(\eta') \theta^{d (1)}
  \theta^{e (1)} \right]\,,
\ee
and  exactly coincides, once more, with the leading order result of Eq. (\ref{LeadStorder3}). This may suggest that the leading lensing terms are correctly resumed by solving (\ref{LensEq}) non-perturbatively.

Let us conclude this section with a comment. The  lens equation as defined by Eq.(\ref{LensEq}) differs by a sign in the last term with respect to the equation used in \cite{Hagstotz:2014qea}. As a consequence, the results given in \cite{Hagstotz:2014qea}  for $\Psi^{(n)}$ to second and  third order  differ from the ones we obtained here. 
Furthermore, as a consequence of this different sign,  the results obtained 
in \cite{Hagstotz:2014qea} for the deformation part of the amplification matrix cannot be seen as  derivatives of the deflection angles with respect to the observed ones.  This is in contrast with our starting definition (\ref{AmpMat}) and with what we have just obtained, but, given the above-mentioned ambiguities in the definition of the amplification matrix, this point needs further study.

%

\section{Conclusions}
\label{Sec6}
\setcounter{equation}{0}

In this paper we have discussed various aspects of the propagation of light-like signals in perturbed cosmological backgrounds making ample use of the so-called geodesic light-cone (GLC) gauge first introduced in \cite{1}.

In practice, we have focussed our attention on the deflection of light rays, showing that the effects of inhomogeneities can be described, in an arbitrary gauge and to the desired level of accuracy, by simply considering (to the same level of accuracy) the coordinate (gauge) transformation that connects the given gauge (in our case the Poisson Gauge (PG)) to those of the GLC, since in this latter the angular coordinates remain constant along null geodesics. We have also verified that, when the deflection is expressed in terms of physical quantities (such as the direction of observation and the observed redshift), exactly the same result is obtained whether one starts from the transformation leading from the PG to the GLC gauge or one proceeds in the opposite direction. 

The explicit calculation was carried out  for the PG  deflection angle in both methods up to the second perturbative order and up to third order for the leading lensing contribution. Given the involved structure of these calculations, finding exact agreement between the two procedures  provides a very non trivial check of their correctedness. 
Our expressions for the PG deflection angle to second and third order ( given in 
Eqs. (\ref{LeadStorder2}), (\ref{Final-theta2-general}) and (\ref{LeadStorder3})), represent one of the main results of this paper.
In that  spirit we have applied the same idea in order to double-check the calculations of the luminosity distance-redshift relation 
$d_L(\theta^a_o, z)$ (in the PG and to full second order) already present in the literature in \cite{3, Fanizza:2013doa} and in 
\cite{Umeh:2012pn, Umeh:2014ana}. 
Such a check is important since the published results are not in full agreement with each other \cite{Marozzi:2014kua}. The result  for $d_L$ (as a function of observer angle and redshift)  reported in 
Appendix \ref{AppA} is in full agreement with the one obtained in  \cite{3, Fanizza:2013doa} using our previous approach \footnote{This however does not exclude the possibility of more conceptual  errors in setting up the calculation in the GLC, or to intrinsic limitations of that gauge, e.g. in the presence of caustics.}.

Concerning the leading-lensing expression for the amplification matrix in the PG at second and third perturbative order  
there appears to be (barring typos) some important discrepancies between our results reported in Sec. \ref{Sec5} and those reported in \cite{Hagstotz:2014qea}. This could alter the conclusions reached in \cite{Hagstotz:2014qea} about  the higher order corrections that  weak lensing could have on the CMB  temperature and polarization anisotropies, but we prefer to leave this interesting issue to further study.

 \vskip 1 cm
 
\section*{Acknowledgements}

We wish to thank Enea Di Dio, Ruth Durrer, Paolo Facchi, Alba Grassi, Fabien Nugier and Saverio Pascazio for useful discussions.
GF and MG are supported in part by MIUR, under grant no. 2012CPPYP7 
(PRIN 2012), and by INFN,  under the program TAsP ({Theoretical Astroparticle Physics}). 
 GF is also supported by the foundation ``Angelo Della Riccia''.
GM is supported by the Marie Curie IEF, Project NeBRiC - ``Non-linear effects and backreaction in classical and quantum cosmology".
GV wishes to acknowledge the hospitality of the KITP Institute in Santa Barbara where part of this work was done.

\appendix

\section{Consistency check for $d_A$ up to second order}
\label{AppA}
\setcounter{equation}{0}

The luminosity-redshift relation $d_L(z,\theta_o^a)$ has been derived in previous papers \cite{3, Fanizza:2013doa}~\footnote{The results of \cite{3, Fanizza:2013doa}
for the luminosity distance-redshift relation are equal to each other, apart from gravitational potential and peculiar velocity terms calculated at the 
observer position. These terms are neglected here.} in the limit of vanishing 
anisotropic stress and up to second order in the Poisson gauge  (see \cite{Marozzi:2014kua} for its generalization  in the presence of anisotropic stress),  starting from the GLC gauge and following the approach 
summarized in Sect. \ref{Sec3}.
On the other hand,  other results about the luminosity-redshift relation have recently  appeared  \cite{Umeh:2012pn, Umeh:2014ana}. These  do not seem to be in complete agreement (see \cite{Marozzi:2014kua}) with those of \cite{3, Fanizza:2013doa}.
We believe that arriving at a commonly accepted expression for the perturbed luminosity-redshift relation is of fundamental importance in view of present (see, for example, \cite{Samushia:2013yga,Delubac:2014aqe}) and future (see \cite{DES,Amendola:2012ys}) high precision observations of cosmic large scale structure.

To this aim we compute here the luminosity-redshift relation, in the limit of vanishing 
anisotropic stress and up to second order in the Poisson gauge, using the approach presented in Sect. \ref{Sec4}, and show that the result exactly coincides with the one obtained in \cite{3, Fanizza:2013doa} (for simplicity, we neglect gravitational potential and velocity terms at the observer position).
This represents, in our opinion,  a highly non-trivial check of the correctness of the result first obtained in \cite{3, Fanizza:2013doa}.

Starting from Eq. (\ref{lumdist}) we first notice that, if we neglect the  gravitational potential and the peculiar velocity contributions at the observer position,  the luminosity distance of a source with general redshift $z$ is just given by (see also \cite{1}):
\be
d_L=(1+z)^2 \frac{(\gamma)^{1/4}}{(\sin \tilde{\theta})^{1/2}}\,.
\label{A1}
\ee
Therefore, we need to evaluate  the full expression of $\gamma_{ab}$ at the second perturbative order.
This can be obtained easily by considering the metric transformation (\ref{5}) written in the form:
\beq
\ga_{ab} = \frac{\pa y^\mu}{\pa \tilde\theta^a}\frac{\pa y^\nu}{\pa \tilde\theta^b} g^{PG}_{\mu\nu} \,,
\eeq
where $\pa_a y^{\mu}$ are exactly the quantities evaluated in Sect.~\ref{Sec4}. Hence, the induced metric on the two-sphere (and consequently the angular/luminosity distance) is given immediately in terms of the GLC angles (i.e. the observed ones) thanks to this new approach.

According to the procedure of Sect. \ref{Sec4} we need, to this purpose, the explicit form of the second order quantities  $\eta^{(2)}$, $\eta^{+(2)}$ and $\theta^{a(2)}$, as well as the corresponding first order expressions.
For $\eta^{+(2)}$ and $\theta^{a(2)}$ we can use the results already presented in Eqs. (\ref{32}) and (\ref{422}), while   for $\eta^{(2)}$ the result 
is still to be computed.

Considering the $g^{00}_{\rm PG}$ component of the coordinate transformation (\ref{5}), we find that $\eta^{(2)}$ must satisfy the following differential equation:
 \bea
\pa_w\eta^{(2)}=&-&a\pa_\tau\eta^{(2)}
-a H \eta^{(2)}
-\frac{1}{2}\phi^{(2)}
+\frac{3}{2}\psi^2
-\frac{1}{2}\left(\pa_w\eta^{(1)}\right)^2
+a H \pa_w\eta^{(1)}\eta^{(1)}\nonumber\\
&+&\pa_w\eta^{(1)}\pa_w\Ep{1}
-\frac{\ga_0^{ab}}{2}\pa_a\eta^{(1)}\pa_b\eta^{(1)}
+\ga_0^{ab}\pa_a\eta^{(1)}\pa_b\Ep{1}\nonumber\\
&+&\pa_w\th{1}{a}\pa_a\eta^{(1)}
+\frac{a^2}{2}\dot{H}\left( \eta^{(1)} \right)^2
-\pa_w\psi\Ep{1}\nonumber\\
&-&\pa_a\psi\,\th{1}{a}
+a H \psi\eta^{(1)}
-\psi\pa_w\eta^{(1)}
-a\pa_\tau\psi\eta^{(1)}\, ,
\label{DEeta2order}
\eea
where we have defined $H=\dot a/a$, and a dot denotes differentiation with respect to $\tau$. In analogy with the first order case, we then find the solution:
\bea
\eta^{(2)}&=&\frac{1}{a(\tau)}\int_{\tau_{in}}^{\tau}d\tau'\left[
-\frac{1}{2}\phi^{(2)}
+\frac{3}{2}\psi^2
-\frac{1}{2}\left(\pa_w\eta^{(1)}\right)^2
+a H \pa_w\eta^{(1)}\eta^{(1)} \right.\nonumber\\
& & \left. +\pa_w\eta^{(1)}\pa_w\Ep{1}
-\frac{\ga_0^{ab}}{2}\pa_a\eta^{(1)}\pa_b\eta^{(1)}
+\ga_0^{ab}\pa_a\eta^{(1)}\pa_b\Ep{1} \right.\nonumber\\
& &\left. +\pa_w\th{1}{a}\pa_a\eta^{(1)}
+\frac{a^2}{2}\dot{H}\left( \eta^{(1)} \right)^2
-\pa_w\psi\Ep{1} \right.\nonumber\\
& &\left. -\pa_a\psi\,\th{1}{a}
+a H \psi\eta^{(1)}
-\psi\pa_w\eta^{(1)}
-a\pa_\tau\psi\eta^{(1)}\right]\left( \tau',\xi(\tau,\tau'),\tilde\theta^a \right)\,,
\eea
which, after some algebraic manipulation,  can be  rewritten in a more useful form as:
\bea
\eta^{(2)}&=&  \frac{1}{a(\tau)} \int_{\tau_{in}}^{\tau}d\tau' \left[-\frac{1}{2}\phi^{(2)}
+\frac{1}{2}\psi^2
-\frac{1}{2}\left(\pa_w\eta^{(1)}\right)^2
-\frac{\ga_0^{ab}}{2}\pa_a\eta^{(1)}\pa_b\eta^{(1)}\right]\left( \tau',\xi(\tau,\tau'),\tilde\theta^a \right)
\nonumber
\\
&&
+\th{1}{a}\pa_a\eta^{(1)}+\left(\eta^{+(1)}-\eta^{(1)}\right) \partial_w \eta^{(1)} - \psi  \eta^{(1)} -\frac{1}{2} a H \left(\eta^{(1)}\right)^2
\,,
\label{eta2orderSimpl}
\eea
where the following relations  have been used:
\bea
\pa_w\,\Ep{1}(\tau',\xi(\tau,\tau'),\tilde\theta^a)&=&
a(\tau')\frac{d}{d\tau'}\,\Ep{1}(\tau',\xi(\tau,\tau'),\tilde\theta^a)+2\,\psi(\tau',\xi(\tau,\tau'),\tilde\theta^a),
\nonumber\\
\pa_w\,\th{1}{a}(\tau',\xi(\tau,\tau'),\tilde\theta^a)&=&
a(\tau')\frac{d}{d\tau'}\,\th{1}{a}(\tau',\xi(\tau,\tau'),\tilde\theta^a)-\ga_0^{ab}\pa_b\,\Ep{1}(\tau',\xi(\tau,\tau'),\tilde\theta^a),
\nonumber\\
a\frac{d}{d\tau'}\,\eta^{(1)}( \tau',\xi(\tau,\tau'),\tilde\theta^a)&=&
-a H \eta^{(1)}(\tau',\xi(\tau,\tau'),\tilde\theta^a)-\psi(\tau',\xi(\tau,\tau'),\tilde\theta^a)\,.
\eea
Similarly, the result for  $\eta^{+(2)}$ given in Eq. (\ref{32}) can also be written in the more compact form as 
\bea
\eta^{+(2)}&=&\int_{\tau}^{\tau_o}\frac{d\tau'}{a(\tau')}\left[\frac{1}{2} \left( \psi^{(2)}+\vp^{(2)} \right)
+\frac{1}{2}\ga_0^{ab}\pa_a\eta^{+(1)}\pa_b \eta^{+(1)}
+2\,\psi^2 -2 \psi \,\pa_w \eta^{+(1)}\right]\nonumber\\
&&
+\eta^{+(1)}\pa_w \eta^{+(1)}+\th{1}{a}\,\pa_a\Ep{1}
-2\,\psi\,\eta^{(1)}.
\label{etaPiuSempl}
\eea

Considering now  the $g^{ab}_{\rm PG}$ component of Eq. (\ref{5}), and extending the previous first-order calculation of $\gamma^{ab (1)}$, we can express $\gamma^{ab (2)}$
in terms of known quantities as:
\bea
\left( \ga^{ab} \right)^{(2)}
&=&\frac{\ga_0^{ab}}{a^2}\left[ 2\,a\,\pa_\tau\psi\,\eta^{(1)}+2\pa_w\psi\Ep{1}+2\th{1}{c}\pa_c\psi+4\psi^2+\psi^{(2)} \right.
-4\,\psi\,\frac{\Ep{1}-\eta^{(1)}}{\ro}
\nonumber
\\
&&
\left. -2\,\frac{\Ep{2}-\eta^{(2)}}{\ro}
+3\,\left(\frac{\Ep{1}-\eta^{(1)}}{\ro}\right)^2
-4\,\psi\,aH\,\eta^{(1)}
+4\,aH\,\eta^{(1)}\,\frac{\Ep{1}-\eta^{(1)}}{\ro} \right.
\nonumber
\\
&&\left.
-2\,aH\,\eta^{(2)}
+a^2\left(H^2-\dot{H}\right)\left( \eta^{(1)} \right)^2\right]
-2\frac{H}{a}\eta^{(1)}\,\th{1}{c}\pa_c\ga_0^{ab}\nonumber
\\
&&
-\frac{2}{a^2}\,\th{1}{c} \pa_c\ga_0^{ab}\frac{\Ep{1}-\eta^{(1)}}{\ro}
+\frac{1}{2}\th{1}{c}\th{1}{d}\frac{\pa_{cd}\ga_0^{ab}}{a^2}
+\th{2}{c}\frac{\pa_c\ga_0^{ab}}{a^2}\nonumber\\
&&+2\psi\,\th{1}{c}\frac{\pa_c\ga_0^{ab}}{a^2}
-\frac{\ga_0^{cb}}{a^2}\pa_c\th{2}{a}
-\frac{\ga_0^{ac}}{a^2}\pa_c\th{2}{b}\nonumber\\
&&+\frac{\ga_0^{ad}\ga_0^{bc}}{a^2}\pa_d\eta^{(1)}\pa_c\Ep{1}
-\frac{\ga_0^{ad}\ga_0^{bc}}{a^2}\pa_d\Ep{1}\pa_c\Ep{1}
+\frac{\ga_0^{ac}\ga_0^{bd}}{a^2}\pa_d\eta^{(1)}\pa_c\Ep{1}\nonumber\\
&&-\frac{2\psi}{a^2}\ga_0^{cb}\pa_c\th{1}{a}
+\frac{\ga_0^{bd}}{a^2}\pa_d\th{1}{c}\pa_c\th{1}{a}
+2\,\frac{H}{a}\ga_0^{bc}\eta^{(1)}\pa_c\th{1}{a}
+2\,\frac{\ga_0^{bc}}{a^2}\frac{\Ep{1}-\eta^{(1)}}{\ro}\pa_c\th{1}{a}\nonumber\\
&&-\frac{\pa_d\ga_0^{bc}}{a^2}\th{1}{d}\pa_c\th{1}{a}
-\frac{2\psi}{a^2}\ga_0^{ca}\pa_c\th{1}{b}
+\frac{\ga_0^{ad}}{a^2}\pa_d\th{1}{c}\pa_c\th{1}{b}
+\frac{\ga_0^{cd}}{a^2}\pa_d\th{1}{a}\pa_c\th{1}{b}\nonumber\\
&&+2\,\frac{H}{a}\ga_0^{ac}\eta^{(1)}\pa_c\th{1}{b}
+2\,\frac{\ga_0^{ac}}{a^2}\frac{\Ep{1}-\eta^{(1)}}{\ro}\pa_c\th{1}{b}
-\frac{\pa_d\ga_0^{ac}}{a^2}\th{1}{d}\pa_c\th{1}{b}\,.
\eea
By applying Eq. (\ref{A1}) we eventually  obtain the following second-order result for the luminosity distance relation:
\bea
d_L(\tau,w,\tilde\theta^a )&=&(1+z)^2\frac{\gamma^{1/4}}{\sin^{1/2}\tilde\theta^1}=d_L^{(0)}+d_L^{(1)}+d_L^{(2)} \equiv \nonumber\\
&=&a\,\left( \Ep{0}-\eta^{(0)} \right)\left\{ 1-\psi+\frac{\Ep{1}-\eta^{(1)}}{\ro}+aH\eta^{(1)}+\frac{1}{2}\nabla_c\,\th{1}{c}\right.\nonumber\\
&&+aH\,\eta^{(2)}
+\frac{a^2}{2}\left(\dot{H}+2H^2\right)\left(\eta^{(1)}\right)^2
-aH\,\psi\,\eta^{(1)}
+\left( aH\,\eta^{(1)}-\psi \right)\frac{\Ep{1}-\eta^{(1)}}{\Ep{0}-\eta^{(0)}}\nonumber\\
&&-a\,\pa_\tau\psi\,\eta^{(1)}
-\pa_w\psi\,\Ep{1}
-\pa_a\psi\,\th{1}{a}
-\frac{\psi^2}{2}
-\frac{\psi^{(2)}}{2}
+\frac{\Ep{2}-\eta^{(2)}}{\Ep{0}-\eta^{(0)}}\nonumber\\
&&-\frac{\ga_0^{ab}}{2}\pa_a\eta^{(1)}\pa_b\Ep{1}
+\frac{\ga_0^{ab}}{4}\pa_a\Ep{1}\pa_b\Ep{1}\nonumber\\
&&\left. -\frac{1}{2}\,\psi\,\nabla_c\,\th{1}{c}
+\frac{1}{2}\,aH\,\eta^{(1)}\,\nabla_c\,\th{1}{c}
+\frac{1}{2}\,\frac{\Ep{1}-\eta^{(1)}}{\Ep{0}-\eta^{(0)}}\,\nabla_c\,\th{1}{c}
+\frac{1}{2}\,\nabla_c\,\th{2}{c}\right.\nonumber\\
&&\left.+\frac{1}{8}\,\left(\nabla_c\,\th{1}{c}\right)^2
-\frac{1}{4}\,\pa_d\th{1}{c}\pa_c\th{1}{d}
-\frac{1}{4}\left(\frac{\theta^{(1)}}{\sin\tilde\theta}\right)^2\right\}\,.
\label{dLAltNoRed}
\eea
(recall that $\nabla_a$ denotes the covariant derivative on the two dimensional sphere with metric $d \sg^2= d \tilde{\theta}^2 +\sin^2 \tilde{\theta} d \tilde{\phi}^2$).

The above result  is written as a function of the GLC coordinates $(\tau, w, \tilde{\theta^a})$. For a clear comparison with the analogous result obtained in  \cite{3, Fanizza:2013doa} we need to express $d_L$ not only in terms of the observation angle $\tilde{\theta}^a_s=\theta_o^a$, but also in terms of the redshift $z_s$, related to the  coordinate $\tau_s$ of the given light source. 
To this purpose we will apply the procedure of Sect. \ref{Sec4}, extending to second order the redshift expansion since, in the case of $d_L$, even the zeroth-order term turns out to be $\tau$-dependent, i.e. $\pa_\tau d_L^{(0)} \not=0$.

By setting $\tau_s=\bar{\tau}_s^{(0)}+\bar{\tau}_s^{(1)}+\bar{\tau}_s^{(2)}$, and by Taylor expanding 
Eq. (\ref{Fiducial-model-GLC-coordinates}) up to second order around 
$\bar{\tau}_s^{(0)}$, we  obtain: 
\bea
1+z_s=\frac{a(\tau_o)}{a(\bar{\tau}^{(0)}_s)}
&=& \frac{a\left(\tau_o\right)}{a(\bar{\tau}^{(0)}_s)}\left\{ 1
- H \bar{\tau}_s^{(1)}+a\left(\Upsilon^{-1}\right)^{(1)}  -H  \bar{\tau}_s^{(2)}+a\pa_\tau\left(\Upsilon^{-1}\right)^{(1)}\bar{\tau}_s^{(1)}\right.
\nonumber\\
&&
\left. \,\,\,\,\,\,\,\,\,\,\,\,\,\,\,\,\,\,\,\,\,\,\,-\frac{1}{2}\left(\dot H-H^2\right)\left( \bar{\tau}_s^{(1)} \right)^2
+a\left( \Upsilon^{-1} \right)^{(2)}\right\}_{\bar{\tau}^{(0)}_s}.
\label{Taylor2order}
\eea
In order for Eq. (\ref{Taylor2order}) to be valid, order by order, $\bar{\tau}_s^{(1)}$ has to satisfy Eq. (\ref{tau1-OR})  while, at second order, we obtain the following equation for $\bar{\tau}_s^{(2)}$:
\beq
\bar{\tau}_s^{(2)}=\frac{1}{H}\left[
a\pa_\tau\left(\Upsilon^{-1}\right)^{(1)}\bar{\tau}_s^{(1)}
-\frac{1}{2}\left(\dot H-H^2\right)\left( \bar{\tau}_s^{(1)} \right)^2
+a\left( \Upsilon^{-1} \right)^{(2)}\right]\,,
\eeq
where the right-hand side has to be evaluated at $\tau=  \bar{\tau}^{(0)}_s$. 
Using 
\be
a\pa_\tau\left(\Upsilon^{-1}\right)^{(1)}
= H\pa_w\Ep{1}-H\pa_w\eta^{(1)}+aH^2\eta^{(1)}+\frac{\pa_w\psi}{a}+\partial_\tau\psi-\frac{\pa^2_w\eta^{(1)}}{a} -a\dot{H}\eta^{(1)}
\,,
\ee
we obtain
\bea
\bar{\tau}_s^{(2)}&=&\frac{1}{H}\left[ H\pa_w\Ep{1}\bar{\tau}_s^{(1)}-H\pa_w\eta^{(1)}\bar{\tau}_s^{(1)}
+aH^2\eta^{(1)}\bar{\tau}_s^{(1)}
+\frac{\pa_w\psi}{a}\bar{\tau}_s^{(1)}+\bar{\tau}_s^{(1)} \partial_\tau \psi 
-\frac{\pa^2_w\eta^{(1)}}{a}\bar{\tau}_s^{(1)}\right.\nonumber\\
&&-\left.a\dot{H}\eta^{(1)}\bar{\tau}_s^{(1)}
-\frac{1}{2}\left(\dot H-H^2\right)\left(\bar{\tau}_s^{(1)} \right)^2
+a\left( \Upsilon^{-1} \right)^{(2)}\right]\,.
\label{tau2-OR}
\eea

To fully evaluate $\tau_s^{(2)}$ from the above equation we thus need  the second order expression of $\Upsilon^{-1}$. Proceeding as in Sect. \ref{Sec4}, and 
considering the coordinates transformation for $g^{+0}_{PG}$, we first obtain:
\bea
\left(\Upsilon^{-1}\right)^{(2)}&=&-\pa_\tau \Ep{2}-\frac{1}{a}\pa_w\Ep{2}
-\pa_\tau\eta^{(2)}
-a\,\pa_\tau\Ep{1}\pa_\tau \eta^{(1)}
-\pa_\tau\Ep{1}\pa_w\eta^{(1)}\nonumber\\
&&-\pa_w\Ep{1}\pa_\tau\eta^{(1)}
+\frac{\ga_0^{ab}}{a}\pa_a\Ep{1}\pa_b\eta^{(1)}
-\frac{1}{\Upsilon^{(1)}}\pa_w\Ep{1}
-\frac{U^{(1)a}}{a}\pa_a\Ep{1}\nonumber\\
&&-\frac{a}{\Upsilon^{(1)}}\pa_\tau\eta^{(1)}
+\frac{4\psi^2}{a}
-\frac{\varphi^{(2)}}{a}
-2\,\pa_\tau\psi\,\eta^{(1)}
-\frac{2}{a}\,\pa_w\psi\,\Ep{1}
-\frac{2}{a}\,\pa_a\psi\,\th{1}{a}\nonumber\\
&&+4\,\psi\frac{\pa_\tau a}{a}\,\eta^{(1)}
-2\,\frac{\pa_\tau a}{a}\,\eta^{(2)}
+2\,\frac{(\pa_\tau a)^2}{a}\,\left(\eta^{(1)}\right)^2
-\frac{\pa_\tau^2 a}{a}\,\left(\eta^{(1)}\right)^2\, .
\eea
By applying the first-order results  for $U^{(1)a}$ and $\left(\Upsilon^{-1}\right)^{(1)}$ given in Eqs. (\ref{27}) and (\ref{29}),
the results  for $\pa_\tau\Ep{1}$, $\pa_\tau\eta^{(1)}$ and $\pa_\tau\th{1}{a}$ 
given in Eqs. (\ref{19}), (\ref{20}) and (\ref{21}),
and by using Eq.(\ref{DEeta2order}), we are then lead to the following 
 result:
\bea
\left(\Upsilon^{-1}\right)^{(2)}&=&\frac{\psi^{(2)}}{2a}
+\frac{3}{2a}\psi^2
-H\eta^{(2)}
-\frac{a}{2}\dot{H}\left( \eta^{(1)} \right)^2
-H\psi\eta^{(1)}
-\frac{\psi}{a}\pa_w\left( \Ep{1}-\eta^{(1)} \right)\nonumber\\
&&+\frac{1}{a}\pa_\eta\psi\,\eta^{(1)}
+\frac{1}{a}\pa_w\psi\,\Ep{1}
+\frac{1}{a}\pa_a\psi\th{1}{a}
-\frac{1}{a}\pa_w\left( \Ep{2}-\eta^{(2)} \right)
+\frac{3}{2a}\left( \pa_w\eta^{(1)} \right)^2\nonumber\\
&&+\frac{1}{a}\left( \pa_w\Ep{1} \right)^2
+H\eta^{(1)}\pa_w\left( \Ep{1}-\eta^{(1)}\right)
-\frac{2}{a}\pa_w\eta^{(1)}\pa_w\Ep{1}\nonumber\\
&&+\frac{1}{a}\pa_w\th{1}{a}\,\pa_a\left( \Ep{1}-\eta^{(1)} \right)
+\frac{\ga_0^{ab}}{2a}\pa_a\eta^{(1)}\pa_b\eta^{(1)}
+\frac{\ga_0^{ab}}{2a}\pa_a\Ep{1}\pa_b\Ep{1}\nonumber\\
&&-\frac{\ga_0^{ab}}{a}\pa_a\eta^{(1)}\pa_b\Ep{1}\, .
\eea
The sought expression for $d_L$ in terms of the observation variables $(\theta_o^a, z_s)$ can now be obtained by Taylor expanding 
Eq.~(\ref{dLAltNoRed}) around $\bar{\tau}^{(0)}_s$ .  Including all contributions up to second order, we can write
\bea
d_L (z_s, w, \theta_o)&=& d_L^{(0)} (\bar{\tau}_s^{(0)} , w, \theta^a_o) + d_L^{(1)}  (\bar{\tau}_s^{(0)} , w, \theta^a_o) 
+\bar{\tau}_s^{(1)}  \partial_\tau \, d_L^{(0)}(\bar{\tau}_s^{(0)} , w, \theta^a_o) 
\nonumber \\
& &  
+  d_L^{(2)} (\bar{\tau}_s^{(0)}, w, \theta^a_o) 
+\bar{\tau}_s^{(2)}  \partial_\tau \, d_L^{(0)}(\bar{\tau}_s^{(0)} , w, \theta^a_o) 
+\frac{1}{2}\left(\bar{\tau}_s^{(1)}\right)^2 \partial^2_\tau \, d_L^{(0)}(\bar{\tau}_s^{(0)} , w, \theta^a_o) 
\nonumber \\
& & 
+\bar{\tau}_s^{(1)}  \partial_\tau \, d_L^{(1)}(\bar{\tau}_s^{(0)} , w, \theta^a_o) \,,
\eea
from which, after a long but straightforward calculation, we can write our final result for $d_L (z_s, w, \theta^a_o)$, as obtained in the new approach of Sect. \ref{Sec4}, as:
\bea
\frac{d_L(z_s,w,\theta_o^a)}{d^{(0)}_L(z_s,w,\theta_o^a)}
&=&1-\psi+\frac{\Ep{1}}{\ro}+\frac{1}{2}\nabla_c\,\th{1}{c}
+\Xi\,\left( \psi-\pa_w\Ep{1}+\pa_w\eta^{(1)} \right)\nonumber\\
&&\!\!\!\!\!\!\!\!\!\!\!\!\!\!\!\!\!\!\!\!\!\!\!\!
+\frac{a^2}{2}\left(\dot{H}+2H^2\right)\left(\eta^{(1)}\right)^2
-aH\,\psi\,\eta^{(1)}
+\left( aH\,\eta^{(1)}-\psi \right)\frac{\Ep{1}-\eta^{(1)}}{\Ep{0}-\eta^{(0)}}\nonumber\\
&&\!\!\!\!\!\!\!\!\!\!\!\!\!\!\!\!\!\!\!\!\!\!\!\!
-a\,\pa_\tau\psi\,\eta^{(1)}
-\pa_w\psi\,\Ep{1}
-\pa_a\psi\,\th{1}{a}
-\frac{\psi^2}{2}
-\frac{\psi^{(2)}}{2}
+\frac{\Ep{2}}{\Ep{0}-\eta^{(0)}}\nonumber\\
&&\!\!\!\!\!\!\!\!\!\!\!\!\!\!\!\!\!\!\!\!\!\!\!\!
-\frac{\ga_0^{ab}}{2}\pa_a\eta^{(1)}\pa_b\Ep{1}
+\frac{\ga_0^{ab}}{4}\pa_a\Ep{1}\pa_b\Ep{1}\nonumber\\
&&\!\!\!\!\!\!\!\!\!\!\!\!\!\!\!\!\!\!\!\!\!\!\!\!
\left. -\frac{1}{2}\,\psi\,\nabla_c\,\th{1}{c}
+\frac{1}{2}\,aH\,\eta^{(1)}\,\nabla_c\,\th{1}{c}
+\frac{1}{2}\,\frac{\Ep{1}-\eta^{(1)}}{\Ep{0}-\eta^{(0)}}\,\nabla_c\,\th{1}{c}
+\frac{1}{2}\,\nabla_c\,\th{2}{c}\right.\nonumber\\
&&\!\!\!\!\!\!\!\!\!\!\!\!\!\!\!\!\!\!\!\!\!\!\!\!
+\frac{1}{8}\,\left(\nabla_c\,\th{1}{c}\right)^2
-\frac{1}{4}\,\pa_d\th{1}{c}\pa_c\th{1}{d}
-\frac{1}{4}\left(\frac{\theta^{(1)}}{\sin\tilde\theta}\right)^2\nonumber\\
&&\!\!\!\!\!\!\!\!\!\!\!\!\!\!\!\!\!\!\!\!\!\!\!\!
+\left[ \Xi+\frac{\dot{H}}{2aH^3\left( \ro \right)} \right]\left(\psi-\pa_w\Ep{1}+\pa_w\eta^{(1)}-aH\eta^{(1)}\right)^2\nonumber\\
&&\!\!\!\!\!\!\!\!\!\!\!\!\!\!\!\!\!\!\!\!\!\!\!\!
+\left[-2\,\psi
+\frac{\Ep{1}-\eta^{(1)}}{\ro}
+\frac{\dot{H}}{H^2\left( \ro \right)}\eta^{(1)}\right.\nonumber\\
&&\!\!\!\!\!\!\!\!\!\!\!\!\!\!\!\!\!\!\!\!\!\!\!\!
+aH\left(\Xi+1\right)\eta^{(1)}
+\Xi\left( \frac{1}{2}\nabla_c\th{1}{c}-2\pa_w\eta^{(1)} +\pa_w\Ep{1}
+\frac{\pa_w\psi}{aH}
- \frac{\pa^2_w\eta^{(1)}}{aH}\right)
\nonumber
\eea
\bea
&&\!\!\!\!\!\!\!\!\!\!\!\!\!\!\!\!\!\!\!\!\!\!\!\!\left.
+\frac{\ga_0^{cd}}{2aH}\nabla_c\pa_d\Ep{1} \right]\left( \psi-\pa_w\Ep{1}+\pa_w\eta^{(1)}-aH\eta^{(1)} \right)\nonumber
\\
&&\!\!\!\!\!\!\!\!\!\!\!\!\!\!\!\!\!\!\!\!\!\!\!\!
+\Xi\left( \frac{\psi^{(2)}}{2}
+\frac{3}{2}\psi^2
-\frac{1}{2}a^2\pa_\tau H\left( \eta^{(1)} \right)^2
-aH\psi\eta^{(1)}
-\psi\pa_w\left( \Ep{1}-\eta^{(1)} \right)\right.\nonumber\\
&&\!\!\!\!\!\!\!\!\!\!\!\!\!\!\!\!\!\!\!\!\!\!\!\!
+\pa_\eta\psi\,\eta^{(1)}
+\pa_w\psi\,\Ep{1}
+\pa_a\psi\th{1}{a}
-\pa_w\left( \Ep{2}-\eta^{(2)} \right)
+\frac{3}{2}\left( \pa_w\eta^{(1)} \right)^2
\nonumber
\\
&&\!\!\!\!\!\!\!\!\!\!\!\!\!\!\!\!\!\!\!\!\!\!\!\!
+\left( \pa_w\Ep{1} \right)^2
+aH\eta^{(1)}\pa_w\left( \Ep{1}-\eta^{(1)}\right)
-2\,\pa_w\eta^{(1)}\pa_w\Ep{1}\nonumber\\
&&\!\!\!\!\!\!\!\!\!\!\!\!\!\!\!\!\!\!\!\!\!\!\!\!
+\pa_w\th{1}{a}\,\pa_a\left( \Ep{1}-\eta^{(1)} \right)
+\frac{\ga_0^{ab}}{2}\pa_a\eta^{(1)}\pa_b\eta^{(1)}
+\frac{\ga_0^{ab}}{2}\pa_a\Ep{1}\pa_b\Ep{1}\nonumber\\
&&\!\!\!\!\!\!\!\!\!\!\!\!\!\!\!\!\!\!\!\!\!\!\!\!
\left.-\ga_0^{ab}\pa_a\eta^{(1)}\pa_b\Ep{1} \right)\,,
\label{FinaldLAlt}
\eea
where we have defined $\Xi=1- [{a H (\ro)}]^{-1}$ and, for simplicity, we have omitted the suffix $s$.

For an explicit comparison of this result with the one obtained in \cite{3, Fanizza:2013doa}, we have to rewrite Eq. (\ref{FinaldLAlt}) using the conformal time $\eta$ and the radial coordinate $r$ 
(or, equivalently, the zero-order light-cone coordinates $\eta^+,\eta^-$) used in \cite{3, Fanizza:2013doa}. Recalling that $d \bar{\tau}_s^{(0)}/a= d\bar{\eta}_s^{(0)}$ (see Sect. \ref{4.3}),  we have
\bea
& & \partial_\tau A(\bar{\tau}_s^{(0)}, w, \theta_o)= (\partial_\eta - \partial_r) A(\eta_s^{(0)}, r_s^{(0)}, \theta^a_o) \label{startE} \,,\\
& & \partial_w A(\bar{\tau}_s^{(0)}, w, \theta_o)=  \partial_r  A(\eta_s^{(0)}, r_s^{(0)}, \theta^a_o) \,,
\eea
for any given  quantity 
$A=A(\bar{\tau}_s^{(0)}, w, \theta_o)$.
On the other hand, considering Eqs. (\ref{partial+-}) and (\ref{PQ}), we also have 
\bea
\Ep{1}&=&-Q \,, \\
\eta^{(1)}&=&-P \,, \\
\pa_w\Ep{1}&=&\psi-\pa_+Q \,,\\
\pa_w\eta^{(1)}&=&-\pa_rP \,.
\eea
Furthermore, we note that the result of \cite{3, Fanizza:2013doa} are expressed in terms of the variables $\tilde{\theta}^{a(1)}$ and
$\tilde{\theta}^{a(2)}$ as given in Eq. (\ref{thetatilde2orderShort}). Therefore, we have to express $\theta^{a(1)}$
and $\theta^{a(2)}$  in terms of such variables.
Using Eq. (\ref{thetatilde2orderShort_Inverted}), and equating Eqs. (\ref{Taylor}) and (\ref{231}), we then obtain
$\theta^{a(1)}=-\tilde{\theta}^{a(1)}$ for the first-order quantities,
while, after some algebraic manipulation, we are lead to the following relation for the second order variables:
\bea
\nabla_a \theta^{a(2)} &=& - \nabla_a \tilde{\theta}^{a(2)} +\partial_b \tilde{\theta}^{a(1)} \partial_a \tilde{\theta}^{b(1)} 
+\frac{1}{(\sin \theta_o)^2} \left(\tilde{\theta}^{(1)}\right)^2+\tilde{\theta}^{a(1)} \nabla_a \left( \nabla_b \tilde{\theta}^{b(1)}\right)
\nonumber 
\\
& & +\partial_a Q \partial_+ \tilde{\theta}^{a(1)}+Q \partial_+ \left( \nabla_a \tilde{\theta}^{a(1)}\right) 
-\frac{1}{2} \nabla_a \left( Q \gamma_0^{ab} \partial_b Q\right)
+\nabla_a \left( P \gamma_0^{ab} \partial_b Q\right). ~~~~~~~~~
\label{finshE}
\eea
Inserting now  Eqs. (\ref{eta2orderSimpl}) and (\ref{etaPiuSempl}) into Eq. (\ref{FinaldLAlt}), and taking into account the results of Eqs. (\ref{startE})-(\ref{finshE}), after a long but straightforward calculation,  we obtain that the luminosity distance-redshift relation obtained with this new method 
exactly coincides with what obtained in \cite{3, Fanizza:2013doa}, both at first and second order.
For example, considering only the first order result and using $\Delta\eta=\eta_o- \bar \eta_s^{(0)}$, we obtain
\bea
\frac{d^{(1)}_L(z_s,w,\theta_o^a)}{d^{(0)}_L(z_s,w,\theta_o^a)}
&=& -\psi_s+\frac{\eta_s^{+(1)}}{\eta_s^{+(0)}-\eta_s^{(0)}}
+\frac{1}{2}\nabla_c\,\th{1}{c}_s
+\Xi\,\left( \psi_s-\pa_w\Ep{1}_s+\pa_w\eta^{(1)}_s \right)\nonumber\\
&=& -\psi_s-\frac{1}{\Delta\eta} Q -\frac{1}{2}\nabla_a \tilde{\theta}^{a(1)}+\left(1-\frac{1}{{\cal H} \Delta\eta}\right)
\left(\partial_+ Q-\partial_r P\right) \nonumber \\
&=& 
-\left(1-\frac{1}{\Hcal_s\Delta \eta}\right) v_{||s}
-\psi_s-
\left(1-\frac{1}{\Hcal_s\Delta \eta}\right)\left[\psi_s
+2 \int_{\eta_s^{(0)}}^{\eta_o} d \eta' \partial_{\eta'}\psi \left(\eta'\right)\right]
\nonumber \\ & &
+\frac{2}{\Delta \eta}  \int_{\eta^{(0)}_s}^{\eta_o} d\eta' \psi(\eta')-\frac{1}{\Delta\eta} \int_{\eta_s^{(0)}}^{\eta_o} d \eta' \,\frac {\eta' - \eta_s^{(0)}}{\eta_o - \eta'} \Delta_2 \psi^I(\eta')
\eea
in full agreement not only with \cite{3, Fanizza:2013doa} but also with \cite{Bonvin:2005ps,Pyne:2003bn}.

We believe, in conclusion, that the above extremely non-trivial check fully confirms the correctness and solidity of our old result (obtained for the first time in \cite{3})  for the luminosity distance-redshift relation up to second order in the Poisson gauge.

\section{Leading lensing contribution to luminosity related observables}
\label{AppB}
\setcounter{equation}{0}

There has been some debate in the literature about which luminosity related observables are free from the leading lensing corrections.
In this Appendix we try to clarify this issue.

We start by summarizing the results obtained in  \cite{3,Fanizza:2013doa} for the luminosity distance-redshift relation $d_L(\theta^a_o, z_s),$ up to second order in scalar perturbations and in the Poisson gauge. We will assume no anisotropic stress\footnote{As seen, anisotropic stress can be taken into account in this limit by simply replacing $\psi$ with $(\psi+\phi)/2$ as discussed in 
 \cite{Marozzi:2014kua}.} 
and keep only the leading lensing terms
 with four angular derivatives. Starting from the result of  \cite{3}, and writing everything explicitly as done  in  \cite{Marozzi:2014kua}, we have, up to  second order:
\bea
\frac{d_L^{\,(1)}}{d_L^{\,(0)}}&=&
-\frac{1}{\Delta\eta}\int_{\eta_s}^{\eta_o}d\eta'\r{\eta'}{\eta_s}\Delta_2 \psi \left( \eta' \right) \,,
\label{dLorder1App}
\\
\frac{d_L^{\,(2)}}{d_L^{\,(0)}}&=&
-\frac{1}{4\Delta\eta}\int_{\eta_s}^{\eta_o}d\eta'\r{\eta'}{\eta_s}\Delta_2\left[ \psi^{(2)}\left( \eta' \right)+\phi^{(2)}\left( \eta' \right) \right]\nonumber\\
& &+\frac{1}{2}\left( \frac{1}{\Delta\eta}\int_{\eta_s}^{\eta_o}d\eta'\r{\eta'}{\eta_s}\Delta_2\psi(\eta') \right)^2\nonumber\\
& &+\frac{2}{\left(\eta_o-\eta_s\right)^2}\int_{\eta_s}^{\eta_o}d\eta'\r{\eta'}{\eta_s}\pa_b\left[\Delta_2 \psi(\eta')\right] \int_{\eta_s}^{\eta_o}d\eta'\r{\eta'}{\eta_s}\bar\ga_0^{ab}\pa_a\psi(\eta')\nonumber\\
& &-2\int_{\eta_s}^{\eta_o}d\eta'\left\{ \ga_0^{ab}\pa_b\left[ \int_{\eta'}^{\eta_o}d\eta''\psi(\eta'') \right]\frac{1}{\eta_o-\eta'}\int_{\eta'}^{\eta_o}d\eta''\r{\eta''}{\eta'}\pa_a\Delta_2\psi(\eta'') \right\}\nonumber\\
&&-\frac{1}{\Delta\eta}\int_{\eta_s}^{\eta_o}d\eta'\r{\eta'}{\eta_s}\Delta_2\left[ \ga_0^{ab}\pa_a\left( \g{\eta'}{\eta''} \right)\pa_b\left( \g{\eta'}{\eta''} \right) \right],
\label{dLorder2App}
\eea
where $\Delta\eta=\eta_o- \bar \eta_s^{(0)}=\bar r_s^{(0)}$, and where $\Delta_2$ is the two-dimensional angular Laplacian referred to the PG angles at the observer position $\theta_o^a= (\theta_o, \phi_o)$, i.e. = $\Delta_2= \pa^2_{\theta_o} + \cot \theta_o \pa_{\theta_o} + (\sin \theta_o)^{-2} \pa^2_{\phi_o}$.

From (\ref{dLorder1App}) and (\ref{dLorder2App}) we can then easily obtain the general solution for any power of $d_L$:
\beq
\frac{(d_L^n)^{\,(1)}}{(d_L^n)^{\,(0)}} = n \frac{d_L^{\,(1)}}{d_L^{\,(0)}}~~;~~ 
~~~~~~~~~~~
\frac{(d_L^n)^{\,(2)}}{(d_L^n)^{\,(0)}} = n \frac{d_L^{\,(2)}}{d_L^{\,(0)}} +\frac{n}{2}\left(n-1\right)\left(\frac{d_L^{\,(1)}}{d_L^{\,(0)}}\right)^2
\,.
\eeq
With a little bit of work the leading second order results can be re-written as a term proportional to the square of 
first order lensing  plus total derivatives terms. More specifically: 
\bea
\frac{\left( d_L^{\,n} \right)^{(2)}}{\left( d_L^{\,n} \right)^{(0)}}&=&\frac{n}{8}\left( n-2 \right)\left( \Delta_2\psiP \right)^2-\frac{n}{4}\frac{1}{\Delta\eta}\int_{\eta_s}^{\eta_o}d\eta'\r{\eta'}{\eta_s}\Delta_2\left[ \psi^{(2)}\left( \eta' \right)+\phi^{(2)}\left( \eta' \right) \right]
\nonumber\\
& &
+\frac{n}{2}\pa_b\left( \Delta_2\psiP\,\bar\ga_0^{ab}\pa_a\psiP \right)
+n\,\pa_b\int_{\eta_s}^{\eta_o}d\eta'
\ga_0^{ab}\pa_a\left( \g{\eta'}{\eta''} \right)\Delta_2\psiP(\eta')
\nonumber\\
& &
-\frac{n}{\Delta\eta}\int_{\eta_s}^{\eta_o}d\eta'\r{\eta'}{\eta_s}\Delta_2\left[ \ga_0^{ab}\pa_a\left( \g{\eta'}{\eta''} \right)\pa_b\left( \g{\eta'}{\eta''} \right) \right],
\eea
where $\psiP(\eta)$ is the lensing potential defined in Eq.(\ref{Lensing-Potential}).

We note that for  $n=2$ the lensing-square term drops out  and we remain with just total derivative terms
(see, also, \cite{Bonvin:2015uha,Bonvin:2015kea}).
As a consequence, if we integrate the inverse flux $\left(d_L/d^{(0)}_L\right)^2$ (also called ``reciprocal magnification" in \cite{Bonvin:2015kea}) around the background sphere we obtain no leading lensing contribution to both first and second order \cite{Bonvin:2015kea}.

The  above results for $d_L$ were already obtained in \cite{BenDayan:2013gc} but  the claim made there  was  that the flux $\Phi \sim d_L^{-2}$, rather than its inverse, has no leading-lensing bias (see also \cite{11a}). This  is in agreement with the argument given
 for the first time by Weinberg in \cite{Weinberg} (see also  \cite{Kibble:2004tm}) that the average of the flux over the celestial sphere has no leading lensing contribution in a universe filled with stochastically isotropic and homogeneous perturbations. 
 
The resolution of this apparent conflict is actually quite simple. In fact, in \cite{BenDayan:2013gc} the definition of angular average includes a proper-area weight on the constant redshift sphere $\Sigma(w_o,z_s)$. 
As a result (see Eq. (2.17) in  \cite{BenDayan:2013gc}):
\beq 
\label{Phitheory}
\langle d_L^{-2} \rangle(w_o,z_s)=(1+z_s)^{-4}\frac{\int dS \frac{d\Omega_o}{dS}}{\int dS}=(1+z_s)^{-4}\frac{\int d\Omega_o}{\int dS}=(1+z_s)^{-4}\frac{4\pi}{\Acal(w_o, z_s) } ~~,
\eeq
with
\beq
\Acal(w_o, z_s) =  \int _{\Sigma(w_o,z_s)} d^2 \tilde{\theta}^a  \sqrt{\ga} ~~
\eeq
where $\langle ... \rangle$ is the light-cone average defined in \cite{1}~\footnote{This light-cone average was introduced by extending to null hypersurfaces the gauge invariant average procedure for space-like domains defined in \cite{Gasperini:2009wp,Gasperini:2009mu,Marozzi:2010qz}.}.
$\Acal(w_o, z_s)$ is the proper area of $\Sigma(w_o,z_s)$ computed with its induced metric $\gamma_{ab}$, and expressed in terms of the coordinates $(w_o,z_s)$ that identify the particular deformed 2-sphere $\Sigma(w_o,z_s)$ on which supernovae at fixed $z$ lie.
Since we have $\sqrt{\gamma}\sim d_L^2$ \cite{Fanizza:2013doa},   
saying that the average of the $d_L^2$ over the directions of observation  has no leading lensing contribution is equivalent to saying that there is no-leading lensing contribution for the area-weighted, angle-averaged 
flux over the past 2-sphere of constant redshift~\footnote{See also  \cite{Kaiser:2015iia} for similar considerations.}.
  
The above result also shows  that the stochastic average introduced in Eq.~(6) of \cite{Bonvin:2015kea} is equivalent, at  the leading lensing level, to the light-cone average used in \cite{BenDayan:2013gc}.
This is because the lensing correction appearing in Eq.~(6) of \cite{Bonvin:2015kea} is equivalent to the lensing correction acting on the induced metric of the deformed 2-sphere $\Sigma(w_o,z_s)$.

  In conclusion, the observable that receives the smallest bias from lensing depends on the precise definition of averaging: we have taken the point of view that the number of sources in a given solid angle is proportional to the proper area subtended by that angle, but ultimately it's physics -- and the way the observations are made -- that should decide which is the mathematical definition that should be compared with the data \cite{Kibble:2004tm}.

\section{Covariance of our expression for $d_L$ }
\label{AppC}
\setcounter{equation}{0}

The luminosity distance $d_L$ has been computed up to second order (and in the Poisson gauge) both without  \cite{3,Fanizza:2013doa}  and with  \cite{Marozzi:2014kua} anisotropic stress. However, the expression for  $d_L$  obtained in \cite{3,Fanizza:2013doa} contains terms which, naively,  do not appear to be covariant  under spatial rotations, a property required for a physical observable like $d_L$.

Considering, for instance,  Eqs. (3.4) and (3.5) of \cite{3} (or the final result given in  Appendix~\ref{AppA}), we find that the suspicious terms appear in the following combination:
\be 
-\frac{1}{2}\nabla_a \tilde{\theta}^{a(2)} 
+\frac{1}{4} \partial_a \tilde{\theta}^{b(1)}
 \partial_b \tilde{\theta}^{a(1)}
 +\frac{1}{4 \sin^2 \theta}\left(\tilde{\theta}^{(1)}\right)^2 \,,
  \label{ANC1}
 \ee
with $\tilde{\theta}^{a(1)}$ and $\tilde{\theta}^{a(2)}$ given in Eq. (\ref{thetatilde2orderShort}).
Hence, the question to address is whether  the expression  in Eq.(\ref{ANC1}) is covariant under rotations. Let us first note that, for such second-order terms,  the angles $\theta^a$ can be safely identified with the observation angles  $\theta_o^a$  without loss of generality.

In order to prove the covariance of Eq.(\ref{ANC1}) let us assume that the combination  
 \be 
 \bar{\theta}^{a(2)}=\tilde{\theta}^{a(2)}-\frac{1}{2} \tilde{\theta}^{d(1)}\partial_d  \tilde{\theta}^{a(1)},
 \label{Useful-Vector}
 \ee
behaves like a true vector. In that case we have that
 \be
 \nabla_a  \bar{\theta}^{a(2)}=\nabla_a \tilde{\theta}^{a(2)}- \frac{1}{2} \partial_a \tilde{\theta}^{b(1)}
 \partial_b \tilde{\theta}^{a(1)} 
 -\frac{1}{2 \sin^2 \theta}\left(\tilde{\theta}^{(1)}\right)^2
  -\frac{1}{2} \tilde{\theta}^{d(1)} \partial_d\left(\nabla_a \tilde{\theta}^{a(1)}\right),
 \ee
and it follows that the sum of the terms of Eq.(\ref{ANC1}) is given by $-(1/2) \nabla_a  \bar{\theta}^{a(2)}$ plus a term which is explicitly covariant (because 
 $\tilde{\theta}^{a(1)}$ is covariant). Hence, to prove that Eq.(\ref{ANC1}) is covariant we only have  to show that our assumption  that 
 $\bar{\theta}^{a(2)}$ is a vector is true.

To this purpose, using the results  (\ref{thetatilde2orderShort}) and  (\ref{thetatilde2orderShort_Inverted})-(\ref{Lensing-Potential}), we can write the ``non trivially covariant" (NTC) part of $\bar{\theta}^{a(2)}$ as follows:
 \be
\left( \bar{\theta}^{a(2)} \right)_{NTC}=-\frac{1}{2}\hat\ga_0^{bc}\pa_c\psiP\pa_b\left( \hat\ga_0^{ad}\pa_d\psiP \right)-2\int_{\eta_s}^{\eta_o}d\eta'\ga_0^{dc}\pa_c\left( \g{\eta'}{\eta''} \right)\pa_d\left( \hat\ga_0^{ab}\pa_b\psiP(\eta') \right),
\ee
which, after an integration by parts, becomes:
 \be
\left(  \bar{\theta}^{a(2)} \right)_{NTC}=\frac{1}{2}\hat\ga_0^{dc}\pa_c\psiP\pa_d\left( \hat\ga_0^{ab}\pa_b\psiP \right)+2\int_{\eta_s}^{\eta_o}d\eta'\hat\ga_0^{dc}\pa_c\psiP(\eta')\pa_d\left[ \ga_0^{ab}\pa_b\left( \g{\eta'}{\eta''} \right) \right]\,.
\ee 
 To check the covariance of the above expression we can explicitly introduce covariant derivatives and rewrite the above equation as
 \bea
\left(  \bar{\theta}^{a(2)} \right)_{NTC}&=&\frac{1}{2}\hat\ga_0^{dc}\pa_c\psiP\nabla_d\left( \hat\ga_0^{ab}\pa_b\psiP \right)
+2\int_{\eta_s}^{\eta_o}d\eta'\hat\ga_0^{dc}\pa_c\psiP(\eta')\nabla_d\left[ \ga_0^{ab}\pa_b\left( \g{\eta'}{\eta''} \right) \right]\nonumber\\
& &\!\!\!\!\!\!\!\!\!\!\!\!\!\!-\frac{1}{2}\hat\ga_0^{dc}\hat\ga_0^{fe}\pa_c\psiP\pa_e\psiP\,\Gamma_{df}^a
-2\int_{\eta_s}^{\eta_o}d\eta'\hat\ga_0^{dc}\pa_c\psiP(\eta')\ga_0^{fe}\pa_e\left( \g{\eta'}{\eta''} \right)\,\Gamma_{df}^a\,,
\label{bar-theta-last-eq}
\eea 
where $\Gamma^a_{bc}$ are the usual Christoffel symbols. On the other hand, 
by integrating by parts, it can be shown that 
\be
\int_{\eta_s}^{\eta_o}d\eta'\hat\ga_0^{dc}\ga_0^{fe}\pa_e\left( \g{\eta'}{\eta''} \right)\pa_c\psiP(\eta')\,\Gamma_{df}^a=-\frac{1}{4}\hat\ga_0^{dc}\hat\ga_0^{ef}\pa_e\psiP\pa_c\psiP\,\Gamma_{df}^a\,.
\ee 
This implies that the second line of Eq.(\ref{bar-theta-last-eq}) is vanishing confirming that $\bar{\theta}^{a(2)}$ is indeed a vector.
    

\end{document}